\documentclass[aps,prb,groupedaddress,preprint,superscriptaddress,showkeys,showpacs]{revtex4-2}
\usepackage{amsfonts}
\usepackage{amsmath}
\usepackage{graphicx}
\usepackage{mathcomp}
\usepackage{stmaryrd}
\usepackage[cal=boondox]{mathalfa}

\usepackage{amsmath}

\usepackage{mathtools}
\usepackage{braket}

\usepackage[colorlinks=true,allcolors=blue]{hyperref}
\usepackage{graphicx}

\usepackage{savesym}
\usepackage{bm}
\savesymbol{iint}
\usepackage{wasysym}
\restoresymbol{WSS}{iint}

\DeclareMathOperator{\Tr}{Tr}

\usepackage{pdfpages} 
\usepackage{pgffor} 

\makeatletter
\AtBeginDocument{\let\LS@rot\@undefined}
\makeatother

\def\supplementfilename{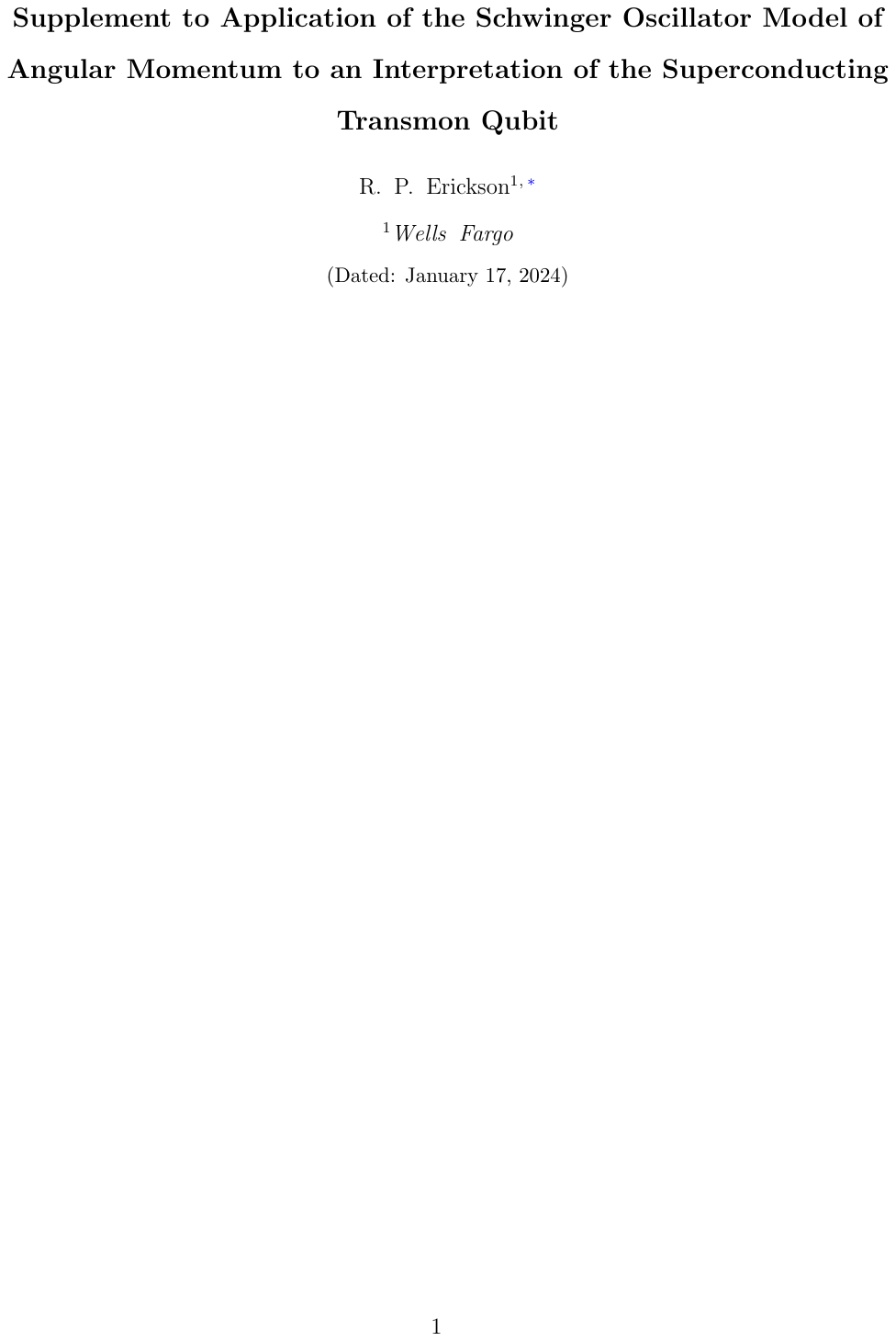}

\pdfximage{\supplementfilename}
\def\numbersupplementpages{\the\pdflastximagepages}

\newif\ifarXiv
\arXivtrue 

\begin{document}

\title{Application of the Schwinger Oscillator Construct of Angular Momentum to an Interpretation of the Superconducting Transmon Qubit}

\author{R. P. Erickson}
\email[]{Electronic address: r.p.erickson2@wellsfargo.com}
\affiliation{Wells Fargo}

\date{\today}

\begin{abstract}
The Schwinger oscillator construct of angular momentum, applied to the superconducting transmon and its transmission-line readout, modeled as capacitvely coupled quantum oscillators, provides a natural and robust description of a qubit. The construct defines quantum-entangled, two-photon states that form an angular-momentum-like basis, with symmetry corresponding to physical conservation of total photon number, with respect to the combined transmon and readout. This basis provides a convenient starting point from which to study error-inducing effects of transmon anharmonicity, surrounding-environment decoherence, and random stray fields on qubit state and gate operations. Employing a Lindblad master equation to model dissipation to the surrounding environment, and incorporating the effect of weak transmon anharmonicity, we present examples of the utility of the construct. First, we calculate the frequency response associated with exciting the ground state to a Rabi resonance with the lowest-lying spin-1/2 moment, via a driving external voltage. Second, we calculate the frequency response between the three lowest two-photon states, within a ladder-type excitation scheme. The generality of the Schwinger angular-momentum construct allows it to be applied to other superconducting charge qubits.
\end{abstract}

\maketitle

\section{Introduction\label{sec:introduction}}
The Jaynes-Cummings model \cite{JaynesCummings1963} was originally conceived to study spontaneous emission and absorption of photons by atoms isolated in a cavity, with the intent to understand stimulated emission of microwaves via coherent amplification in masers \cite{GordonZeigerTownes1955}. The model has since been adapted to a superconducting transmon, for example, to describe a qubit defined by the two lowest-lying (ground and excited) photon states, assuming these states are sufficiently isolated from higher-lying states \cite{Blais2004,*Koch2007,*Schreier2008}. In this circuit quantum electrodynamics (CQED) model, the qubit (the simulated atom) is capacitively or inductively coupled to a transmission line (the resonator cavity), modeled as a linear quantum oscillator, and driven by an externally applied time-varying field (the interaction). The full quantum wave function of the open transmon-resonator system is estimable in terms of a basis of two-photon equilibrium states constructed from products of transmon and resonator equilibrium states. Usually a transformation of the time-dependent Hamiltonian via the interaction picture is made \cite{Sakurai1985-InteractionPicture}, from which a rotating-wave approximation of the driving term can be inferred \cite{BlochAndSiegert}, followed by a transformation of slow-rotation terms to a time-independent Hamiltonian, when possible \footnote{See \cite{BlochAndSiegert} for a discussion of rotating-wave approximation applicability and limitations.}. Regardless of the set of techniques employed in its solution, the Jaynes-Cummings model has proven invaluable to the understanding of superconducting charge-qubit dynamical behavior. 

However, the application of the Jaynes-Cummings model in this simplest form has drawbacks. For example, as mentioned, the two lowest-lying states must be in isolation from those above, which is not always strictly viable, requiring multilevel extensions to the model \cite{Koch2007}. Examples of such multilevel cQED calculations can be found in the literature \cite{Fink_2009,*Kockum_2013}. Also, the strength and relative positions of energy levels, between qubit and resonator, must be engineered to affect as little change on the qubit state during readout as possible, which is counter to the unavoidable intrinsic photon entanglement of the combined transmon and resonator. These issues complicate the definition of the qubit and introduce sources of error in measurement of its coherent state, as well as gate operations applied to it. An alternative is to exploit the tandem design of transmon and resonator, by leveraging the Schwinger oscillator model of angular momentum to construct a more robust definition of the qubit \cite{osti_4389568,*Bidenharn1965}, one rooted in the intrinsic entanglement of the photons, regardless of states excited during operation. This does not change the physics of the device. Instead, it offers a more complete perspective of state entanglement, within the combined transmon and resonator, which can aid device design with respect to quantum decoherence and sources of gate error.

Many years ago J. Schwinger recognized the equivalence between the Lie algebra of annihilation and creation operators, of two adjoining linear quantum oscillators, and the algebra of angular momentum \cite{osti_4389568}. This inference has utility in elementary particle physics as a means of demonstrating the property of nuclear isospin, and how it emerges from the entanglement of up and down quarks \footnote{An excellent description of the Schwinger oscillator model of angular momentum can be found in pages 217-223 of the text of J. J. Sakurai \cite{Sakurai1985-InteractionPicture}}. This same idea can be applied to the transmon and resonator tandem as well, where both are modeled as full quantum oscillators, allowing for a robust definition of a metric of entanglement, wherein the qubit is a naturally emergent property of the transmon and resonator. However, like nuclear isospin, a qubit defined in this way is not a physical observable of angular momentum, though it possesses the same properties of wave-function construction, operator commutation, and operator addition. Moreover, it is not relegated to a specific spin quantum number, such as spin 1/2, since it is not mapped to a specific state excitation.

Most importantly, the Schwinger construct introduces angular-momentum-like symmetry to the combined transmon and resonator, based on an underlying conservation of total photon number, $N$. Specifically, as we shall see, $N$ dictates the spin quantum number, $S$, where $S=N/2$. This creates an analogy with the simulated atom, as in the original intent of the Jaynes-Cummings model, defining specific angular-momentum-based selection rules that govern allowed state transitions, assuming a fixed value of $N$. In this way, the angular momentum symmetry becomes a baseline from which to study sources of measurement error, particularly for the transmon, for which anharmonicity is relatively weak compared to other charge qubits. From the point of view of the Schwinger construct, sources of error, such as transmon anharmonicity or stray stochastic fields, are symmetry breaking mechanisms that respectively can cause $N$ to be perturbed from an integer value or cause it to fluctuate randomly, manifestly as noise-induced error.

In this study we define the Schwinger qubit construct and, as an example application, apply it to an analysis of transmon excitation, introducing a Lindblad master equation to address dissipation with the surrounding environment \cite{Breuer2002}. First, we calculate the Rabi resonance between ground state and first (spin-1/2) excited state as a function of the strength of capacitive coupling between transmon and resonator. Second, among the three lowest energy states of combined transmon and resonator, we calculate the two Rabi resonances of a ladder-type frequency-response scheme, showing the viability of this scheme to indirectly excite the third (spin-1) state. Last, we conclude with some final remarks regarding symmetry breaking and the exposure of the superconducting transmon to one-half noise.

\section{Theory\label{sec:model}}
We first present the model Hamiltonian of the combined transmon and resonator. We then define the qubit of this Hamiltonian via the Schwinger oscillator model of angular momentum. Last, we introduce the voltage driving term and the Lindblad master equation, whose solution is used to calculate the expectation value of the qubit. Additional details can be found in the accompanying Supplemental Material \cite{Supplement2023}.

\subsection{The Transmon and Resonator Hamiltonian}
We model a superconducting transmon of charging energy $E_C$ and Josephson energy $E_J$ as an anharmonic quantum oscillator, where the cosine of the superconducting phase is expanded as a Taylor series, in a Duffing approximation, and the offset charge number is assumed negligible or otherwise removable from the resulting Hamiltonian. Denoting the transmon oscillator ladder by index $-$, its fundamental frequency is $\omega_-=\sqrt{8 E_C E_J}/\hbar$. Similarly, representing the resonator as a linear quantum oscillator of self inductance $L$ and capacitance $C$, and denoting its ladder by index $+$, we have a fundamental frequency of $\omega_+=1/\sqrt{L C}$. With the two oscillators capacitively coupled via parameter $g$, the second-quantized Hamiltonian can be expressed as
\begin{equation}	\label{eq:hamiltonian}
\mathcal{H}_o = \sum_{\mu=\pm} \hbar \omega_\mu \left( a^\dagger_\mu a_\mu + \frac{1}{2} \right)
+ \frac{1}{4} i \hbar \tilde{g} \left( a^\dagger_+ + a_+ \right) \left( a^\dagger_- - a_- \right) - \frac{1}{12} E_C { \left( a^\dagger_- + a_- \right) }^4 ,
\end{equation}
where $\tilde{g}=g\sqrt{\hbar \omega_-/E_C}$ and $a_\mu$, $a^\dagger_\mu$ are annihilation and creation operators, respectively, with commutation relations $[a_\mu,a^\dagger_{\mu'}]=\delta_{\mu,\mu'}$, $[a_\mu,a_{\mu'}]=0$, and $[a^\dagger_\mu,a^\dagger_{\mu'}]=0$. 

Applying a canonical transformation to (\ref{eq:hamiltonian}) to diagonalize linear terms, we arrive at
\begin{equation}	\label{eq:hamiltonian-canonical}
\mathcal{H}_o = \sum_{\sigma\in\left\{ \uparrow, \downarrow \right\}} \hbar \omega_\sigma \left( a^\dagger_\sigma a_\sigma + \frac{1}{2} \right)
- \frac{1}{12} E_c { \left[
\xi_{\uparrow,-} \left( a^\dagger_\uparrow - a_\uparrow \right) - i \xi_{\downarrow,-} \left( a^\dagger_\downarrow + a_\downarrow \right) 
\right] }^4 ,
\end{equation}
where $\uparrow$, $\downarrow$ are indexes denoting the canonical ladders of fundamental frequency
\begin{equation}	\label{eq:frequencies-canonical}
\omega_\sigma = \sqrt{ \frac{1}{2} \left[ \omega^2_+ + \omega^2_- + \sigma \sqrt{ {\left( \omega^2_+ - \omega^2_- \right)}^2 + \tilde{g}^2 \, \omega_+ \omega_- } \right] } ,
\end{equation}
and $\xi_{\sigma,\mu}=\sqrt{(\omega_\sigma/\omega_\mu)(\omega_{\bar{\mu}}^2 - \omega_\sigma^2)/(\omega_{\bar{\sigma}}^2 - \omega_\sigma^2)}$. In this notation, the numerical value of $\sigma$ is $+1$ ($-1$) when index $\sigma=\uparrow$ ($\sigma=\downarrow$), with additional index notations $\bar{\mu}=-\mu$ and $\bar{\sigma}=\downarrow$ ($\bar{\sigma}=\uparrow$) when $\sigma=\uparrow$ ($\sigma=\downarrow$). Note that as $g$, or equivalently $\tilde{g}$, increases then $\omega_\downarrow\rightarrow 0$, as in Fig.~\ref{fig1}(a), indicative of instability; hence, we shall always assume the transmon is weakly coupled to the resonator. In this weak-coupling limit, the $\uparrow$ ($\downarrow$) canonical ladder is most strongly identifiable with the $+$ ($-$) original ladder, i.e., $\omega_\uparrow\cong\omega_+ + \tilde{g}^2 \omega_-/[8(\omega_+^2-\omega_-^2)]$ and $\omega_\downarrow\cong\omega_- - \tilde{g}^2 \omega_+/[8(\omega_+^2-\omega_-^2)]$.

\begin{table}[!htbp]
\caption{\label{table1} Model parameters of transmon and resonator as utilized in the text.}
\begin{ruledtabular}
\begin{tabular}{c|c|c|c||c|c|c|c||c}
\multicolumn{4}{c||}{Resonator} & \multicolumn{4}{c||}{Transmon} &\\
\hline
$L$~(pH) & $C$~(nF) & $\gamma_+'$~(kHz) & $\gamma_+$~(kHz) & $E_C$~($\mu$eV) & $E_J$~($\mu$eV) & $\gamma_-'$~(kHz) & $\gamma_-$~(kHz) & $g$~(MHz) \\
\hline
10.0 & 1.0 & 100.0 & 10.0 & 0.165 & 8.24 & 10.0 & 1.0 & 5.0
\end{tabular} 
\end{ruledtabular} 
\end{table}

\subsection{The Schwinger Oscillator Model of Angular Momentum}
Slow-rotation constituents of $\mathcal{H}_o$ can be expressed in terms of number operators $n_\sigma = a^\dagger_\sigma a_\sigma$ and Schwinger angular momentum components, the latter of which are defined as \footnote{Note that we omit the use of $\hbar$ in the definition of the qubit spin components of (\ref{eq:spin-plus-minus}) and (\ref{eq:spin-cartesian}) since the qubit is not an angular momentum.}
\begin{equation}	\label{eq:spin-plus-minus}
S_+ = a^\dagger_\uparrow a_\downarrow , \quad
S_- = a^\dagger_\downarrow a_\uparrow ,
\end{equation}
\begin{equation}	\label{eq:spin-cartesian}
S_x = \frac{1}{2} \left( S_+ + S_- \right) , \quad
S_y = \frac{1}{2i} \left( S_+ - S_- \right) , \quad
S_z = \frac{1}{2} \left( a^\dagger_\uparrow a_\uparrow - a^\dagger_\downarrow a_\downarrow \right) .
\end{equation}
The length $S=N/2$, proportional to total number operator $N$, is a good quantum number in the absence of nonlinear (anharmonic) terms in $\mathcal{H}_o$. The Cartesian operator components $S_x$, $S_y$, and $S_z$, whose expectation values define the qubit, satisfy the usual commutation relations of angular momentum, i.e., $[S_j,S_k]=i \epsilon_{j,k,l} \, S_l$, where $j,k,l\in\left\{ x,y,z \right\}$ and $\epsilon_{j,k,l}$ is a Levi-Civita coefficient. This result follows from the commutation relations of $a_\sigma$ and $a^\dagger_\sigma$. Similarly, via an inverse canonical transformation, the Schwinger angular momentum components also can be expressed in terms of $a_\mu$ and $a^\dagger_\mu$, with preservation of $[S_i,S_j]=i \epsilon_{i,j,k} \, S_k$.

In the linear limit of (\ref{eq:hamiltonian-canonical}), eigenstates are two-particle states $\ket{n_\uparrow}\otimes\ket{n_\downarrow}$, where $a_\sigma \ket{n_\sigma}=\sqrt{n_\sigma}\ket{n_\sigma -1}$ and $a^\dagger_\sigma \ket{n_\sigma}=\sqrt{n_\sigma +1}\ket{n_\sigma +1}$, with eigenvalues $E_{n_\uparrow,n_\downarrow}=\sum_\sigma \hbar\omega_\sigma \left( n_\sigma + 1/2 \right)$. In this limit, angular momentum states are identical to two-particle energy eigenstates, but with reordered indexing, such that an angular momentum state is of the form
\begin{equation}	\label{eq:angularMomentumState}
\ket{S,m_S} = \frac{{a^\dagger_\uparrow}^{S+m_S} \ket{0} \otimes {a^\dagger_\downarrow}^{S-m_S} \ket{0} }{\sqrt{\left(S+m_S\right)! \left(S-m_S\right)!}} ,
\end{equation}
where $\ket{0,0}=\ket{0}\otimes\ket{0}$ is the vacuum state, length $S=\left( n_\uparrow + n_\downarrow \right) / 2=0,1/2,1,\dots$, and quantum number $m_S=\left( n_\uparrow - n_\downarrow \right) / 2=-S, -S+1, \dots, S-1,S$. Thus, $\ket{S,m_S}$ is an eigenstate of the linear Hamiltonian, with eigenvalue $\hbar\xi_{S,m_S}$, where 
\begin{equation}	\label{eq:spin-frequency}
\xi_{S,m_S} = \left( \omega_\uparrow + \omega_\downarrow \right) \left( S + \frac{1}{2} \right) + \left( \omega_\uparrow - \omega_\downarrow \right) m_S .
\end{equation}
Since operators $S_\pm$ of (\ref{eq:spin-plus-minus}) are legitimate angular momentum operators, we also have the identities
\begin{equation}
S_\pm \ket{S,m_S} = \sqrt{\left( S \mp m_S \right) \left( S \pm m_S + 1 \right)} \ket{S,m_S \pm 1} ,
\end{equation}
as the reader can verify. Note that states $\ket{S,m_S}$ of (\ref{eq:angularMomentumState}) are symmetric since the particles of the combined oscillators are bosons. 

\begin{figure}[!htbp]
\center
\includegraphics[width=240pt, height=400pt]{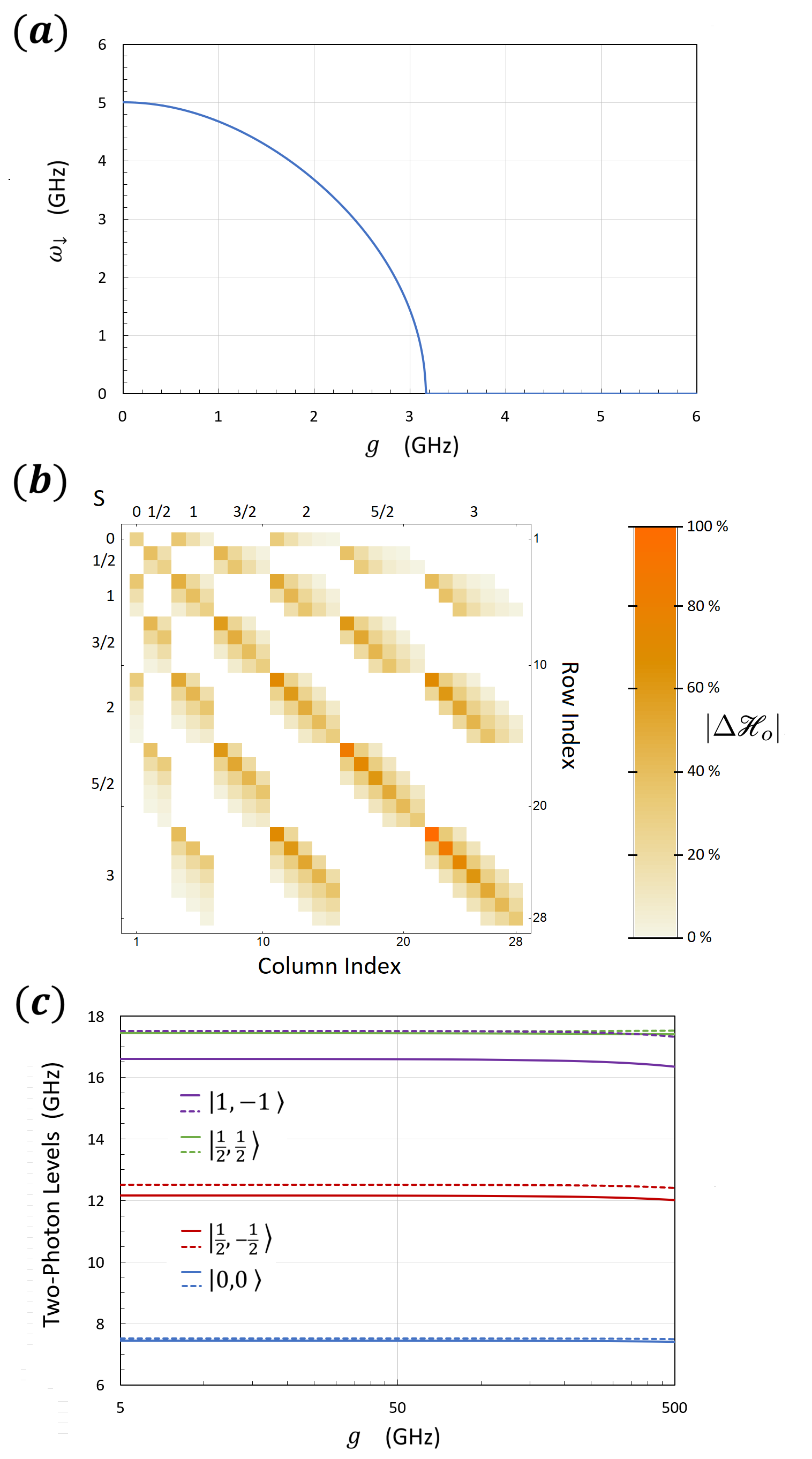}
\caption{\label{fig1} (a) Canonical ladder frequency $\omega_\downarrow$ plotted as a function of capacitive coupling strength $g$, showing how modes of the $\downarrow$ ladder go soft with increasing $g$, at $g_c=3.17$~GHz, indicative of an instability. (b) Magnitude of nonlinear matrix elements of time-independent Hamiltonian, $\left| \Delta \mathcal{H}_o \right|$, plotted as a heat map, on a relative scale, row index vs column index, with labels indicating blocks of constant $S$. White regions indicate zero-valued matrix elements, consistent with forbidden state transitions, as discussed in the text. (c) Two-photon energy levels plotted in units of GHz as a function of $g$. As discussed in the text, dashed curves are calculated without anharmonicity while solid curves include anharmonicity via second-order perturbation theory. All figures were made utilizing the parameters stated in Table~\ref{table1}, as applicable.}
\end{figure}

An important point to note is that the Schwinger construct introduces selection rules via the angular-momentum symmetry, associated with constant $N$. To see this, consider defining the nonlinear terms of $\mathcal{H}_o$ of (\ref{eq:hamiltonian-canonical}) as
\begin{equation}	\label{eq:hamiltonian-nonlinear-terms}
\Delta \mathcal{H}_o = -\frac{1}{12} E_c { \left[
\xi_{\uparrow,-} \left( a^\dagger_\uparrow - a_\uparrow \right) - i \xi_{\downarrow,-} \left( a^\dagger_\downarrow + a_\downarrow \right) 
\right] }^4 .
\end{equation}
Let two-photon eigenstates of $\mathcal{H}_o$ be denoted by $\ket{\psi_k}$, with indexes $k=1,2,\dots$, such that $\mathcal{H}_o \ket{\psi_k} = \hbar \epsilon_k \ket{\psi_k}$, where $\hbar \epsilon_k$ is an eigenstate energy. For the transmon, where the anharmonicity is weak, i.e., $E_C\ll 12\hbar\omega_\downarrow$, we can use standard Rayleigh-Schr\"{o}dinger time-independent perturbation theory \footnote{See, for example, pages 289-293 of \cite{Sakurai1985-InteractionPicture}.} to evaluate $\ket{\psi_k}$ and $\epsilon_k$, using the states $\ket{S,m_S}$ of (\ref{eq:angularMomentumState}) as a basis. 

In Fig.~\ref{fig1}(b) we plot the relative magnitude of the matrix elements $\braket{\psi_k|\Delta \mathcal{H}_o|\psi_{k'}}$, for $k,k'=1,2,...,\left(S+1\right)\left(2S+1\right)$, as a heat map, evaluating $\bra{\psi_k}$ and $\ket{\psi_{k'}}$ to second order in the perturbation expansion, using a subset of states $\ket{S,m_S}$, up to $S=3$. The plot shows the relative magnitude of each element by row ($k$) and column ($k'$), where white space denotes a magnitude of zero. Rows $k$ and columns $k'$ are indicated on the right and bottom edges of the heat map, respectively. The left and top edges of the heat map are labeled with the spin quantum numbers most closely identifiable (from the perturbation theory) with indexes $k$ and $k'$. The elements $\braket{\psi_k|\Delta \mathcal{H}_o|\psi_{k'}}$ are shown to form well-defined blocks of non-zero sub matrices according to the value of spin $S$, with large non-zero magnitudes aligned in bands parallel to the matrix diagonal.

In zero order of the perturbation expansion, only blocks of matrices along the diagonal---rows and columns of the same value of $S$---would be non-zero. In second order of the perturbation theory, however, we see from the figure that non-zero blocks extend to bands on either side of the non-zero diagonal blocks, but only for sub matrices where the row-value of $S$ and the column-value of $S$ differ by an integer value. This indicates that, for the superconducting transmon, atom-like selection rules apply, even in the presence of weak anharmonicity. In this specific situation, allowed state transitions are associated with integer change in the value of spin; half-integer changes in spin are essentially forbidden. Hence, since $S=N/2$, this means that the transmon tends to be robust against changes in $N$ that involve an odd number of photons, such as those associated with random one-half noise events. In contrast, changes in $N$ by an even number of photons include events such as pairwise adiabatic, elastic collisions that conserve linear momentum, as when the transmon-resonator system is subjected to a coherent applied voltage.

Lastly, we show the form of energy levels $\hbar \epsilon_k$ calculated from the second-order perturbation theory. Specifically, in Fig.~\ref{fig1}(c), we plot the two-photon $\epsilon_k$, measured in GHz, as a function of capacitive coupling strength $g$, also measured in GHZ, for the first four energy levels. The dashed curves are the zero-order, unperturbed values---the same as (\ref{eq:spin-frequency})---plotted for comparison with the solid-curve, second-order estimates of $\epsilon_k$. The curves are labeled by the kets of their zero-order correspondences. Note that the purple solid curve ($\ket{1,-1}$) is lower in value than that of the green solid curve ($\ket{\frac{1}{2},\frac{1}{2}}$), a juxtaposition indicative of an avoided crossing, resulting from the anharmonicity.

\subsection{The Driving Voltage and Lindblad Master Equation}
We model the interaction of the qubit with an externally applied field by introducing a time-dependent energy term $\mathcal{U}(t)=-q_+ \, V(t)$, where $V(t)$ is the driving voltage and $q_+$ is the charge of the resonator. As derived in Section~III of the Supplemental Material \cite{Supplement2023}, $\mathcal{U}(t)$ can be expressed in second-quantized form, within the canonical representation, as
\begin{equation}	\label{eq:U-t}
\mathcal{U}(t) = -\hbar \Theta \, V(t) ,
\end{equation}
\begin{equation}	\label{eq:voltage-coefficient}
\Theta = v_\uparrow \left( a^\dagger_\uparrow + a_\uparrow \right) + i v_\downarrow \left( a^\dagger_\downarrow - a_\downarrow \right) ; \quad
v_\sigma = \frac{1}{\sqrt{2\hbar Z}} \left( \frac{\omega_+}{\omega_\sigma} \right) \xi_{\sigma,+} .
\end{equation}
In this way, the Hamiltonian is modified to an open time-dependent form $\mathcal{H}(t)=\mathcal{H}_o+\mathcal{U}(t)$. 

Additionally, to complete the description of the open system, we model the interaction of transmon and resonator with the surrounding environment via a Lindblad master equation \cite{Breuer2002}, where $\gamma_\mu'$ and $\gamma_\mu$ are diffusion and dissipation rates, respectively, between environment and resonator ($\mu=+$), and environment and transmon ($\mu=-$). As alluded to earlier, in Fig.~\ref{fig1}(a), we assume $|\tilde{g}|\ll 4\omega_\mu$, such that, to first approximation, we can neglect diffusion and dissipation directly between transmon and resonator. The canonical form of the Lindblad master equation can be expressed as
\begin{multline}	\label{eq:lindbladMasterEquation}
\frac{d \rho(t)}{d t} = \frac{1}{i\hbar} \left[ \mathcal{H}(t), \rho(t) \right] \\
+ \frac{1}{2} \sum_{\sigma,\sigma'\in\left\{\uparrow,\downarrow\right\}} \sum_{s,s'=\pm}
\gamma_{\sigma,s;\sigma',s'} \left[ 2 A_{\sigma,s} \rho(t) A^\dagger_{\sigma',s'} - A^\dagger_{\sigma',s'} A_{\sigma,s} \rho(t) 
- \rho(t) A^\dagger_{\sigma',s'} A_{\sigma,s} \right] ,
\end{multline}
where $\rho(t)$ is a density operator, $A_{\sigma,+}=a_\sigma$ and $A_{\sigma,-}=a^\dagger_\sigma$ are jump operators, and
\begin{multline}	\label{eq:lindbladCoefficient}
\gamma_{\sigma,s;\sigma',s'} = \frac{1}{4} \sum_{\mu,m=\pm} \gamma_{\mu,m} \, \xi_{\sigma,\mu} \, \xi_{\sigma',\mu}
\left( \frac{ \omega_\mu + m s \, \omega_\sigma }{ \omega_\sigma } \right) \left( \frac{ \omega_\mu + m s' \, \omega_{\sigma'} }{ \omega_{\sigma'} } \right) \\
\times e^{i\pi \left[ \left( \sigma - \sigma' \right) m \left( 1 - \mu \right) - s \left( 1 - \sigma \right) + s' \left( 1 - \sigma' \right) \right] / 4} ,
\end{multline}
with $\gamma_{\mu,m} = \left( \gamma_\mu' + \gamma_\mu \right) \delta_{m=+} + \left( \gamma_\mu' - \gamma_\mu \right) \delta_{m=-}$.

In practice, we solve (\ref{eq:lindbladMasterEquation}) in component form by defining matrix elements of $\rho(t)$ using the energy states $\ket{\psi_k}$ of $\mathcal{H}_o$ as a basis, calculated from the second-order perturbation theory described earlier. We then have matrix elements $\rho_{k,k'}(t)=\braket{\psi_k|\rho(t)|\psi_{k'}}$, as well as $\Theta_{k,k'}=\braket{\psi_k|\Theta|\psi_{k'}}$ for $\Theta$ of (\ref{eq:voltage-coefficient}), and $A_{k,k'}(\sigma,s)=\braket{\psi_k|A_{\sigma,s}|\psi_{k'}}$ for jump operator $A_{\sigma,s}$. We also truncate the infinite set of coupled differential equations to only those that interact strongly with $V(t)$, labeling this finite set of states by $\mathcal{E}$. In this way, we set $\rho_{k,k'}(t)=0$ unless $k,k'\in\mathcal{E}$, such that the component form of (\ref{eq:lindbladMasterEquation}) is approximated as
\begin{equation}	\label{eq:lindbladMasterEquation-subset}
\frac{d \rho_{k,k'}(t)}{d t} \cong -i \left( \epsilon_k - \epsilon_{k'} \right) \rho_{k,k'}(t) 
+ \sum_{l,l'\in\mathcal{E}} \left[ i \left( \Theta_{k,l} \, \delta_{k',l'} - \Theta_{l',k'} \, \delta_{k,l} \right) V(t) 
+ \Lambda^{(l,l')}_{k,k'}\right] \rho_{l,l'}(t) ,
\end{equation}
where $k,k'\in\mathcal{E}$ and we have defined coefficients
\begin{multline}	\label{eq:lindbladMasterEquation-coefficient}
\Lambda^{(l,l')}_{k,k'} = 
\frac{1}{2} \sum_{\sigma,\sigma'\in\left\{\uparrow,\downarrow\right\}} \sum_{s,s'=\pm}
\gamma_{\sigma,s;\sigma',s'} \Bigg\{ 2 A_{k,l}(\sigma,s) \, A_{k',l'}(\sigma',s')^* \\
- \sum_{k''\in\mathcal{E}} \left[ 
A_{k'',k}(\sigma',s')^* \, A_{k'',l}(\sigma,s) \, \delta_{k',l'} + A_{k'',l'}(\sigma',s')^* \, A_{k'',k'}(\sigma,s) \, \delta_{k,l} \right] \Bigg\} ,
\end{multline}
indicative of the strength of interaction with the surroundings. Similar to $\rho_{k,k'}(t)$, a matrix element of the expectation value of a Schwinger spin component can be expressed as $S^{(\alpha)}_{k,k'}=\braket{\psi_{k}|S_\alpha|\psi_{k'}}$, where $\alpha\in\left\{x,y,z\right\}$, such that $\braket{{\bf S}(t)}=\Tr\left[{\bf S}\rho(t) \right]$ can be approximated as
\begin{equation}	\label{eq:qubit-expectation-value}
\braket{{\bf S}(t)} \cong \sum_{\alpha\in\left\{x,y,z\right\}} \sum_{k,k'\in\mathcal{E}} S^{(\alpha)}_{k,k'} \, \rho_{k',k}(t) \, \hat{\alpha} ,
\end{equation}
where $\hat{\alpha}$ represents a unit vector of the abstract Cartesian coordinate system.

\section{Results\label{sec:results}}
In the following calculations we use the parameters listed in Table~\ref{table1}, unless otherwise stated. For concreteness, regarding the resonator, we take the self inductance to be $L=10.0$~pH and the capacitance to be $C=1.0$~nF, such that $\omega_+=1/\sqrt{L C}=10.0$~GHz. For the transmon we assume $E_J/E_C=50$ and take $\omega_-=5.0$~GHz, which requires $E_C=0.165~\mu$eV and $E_J=8.24~\mu$eV. Also, we set the diffusion and dissipation rates of (\ref{eq:lindbladMasterEquation}) to values that correspond to coherence times in the range of $100~\mu$s, allowing the resonator to be more interactive with the surrounding environment than the transmon, with $\gamma_+'=100$~kHz, $\gamma_+=10.0$~kHz and $\gamma_-=10.0$~kHz, $\gamma_-=1.0$~kHz.

We first present results for excitation of the ground state to the lowest-lying excited state of the transmon-resonator system, corresponding to an induced qubit of spin-1/2, near the Rabi resonance frequency, using a single-tone applied voltage. This calculation uses the two lowest-lying energy levels plotted in Fig.~\ref{fig1}(c). In this analysis we derive the coherence time of the qubit, which we set, via Table~\ref{table1}, to correspond to about $100~\mu$s. Also, we extend the two-state calculation to consider the frequency response as a function of increasing $g$. In a second set of results, we apply a two-tone voltage and explore the excitation of the ground state to the two lowest-lying states, of spin-1/2 and spin-1, respectively, within a Rabi resonance calculation corresponding to a three-state, ladder-type frequency-response scheme. Again, we make use of Fig.~\ref{fig1}(c), but in this case using the three lowest-lying energy levels. This three-state resonance calculation demonstrates the excitation of the Schwinger qubit to a superposition that includes the spin-1/2 and spin-1 states.

\subsection{A Single-Tone Two-State Rabi Resonance Calculation}
We first considered the solution of (\ref{eq:lindbladMasterEquation-subset}) for a Rabi resonance between the two lowest-lying states, for which $\mathcal{E}=\left\{ 1, 2 \right\}$. In this case, a single-tone voltage, $V(t)=V_o \, \sin{\Omega t}$, with $V_o=1.0$~nV, was applied for values of pump frequency $\Omega\approx\epsilon_{2,1}$, where $\epsilon_{2,1}=\epsilon_2-\epsilon_1$. Since transmon anharmonicity is weak, we made a first approximation of coefficients $\Theta_{k,k'}$ and $A_{k,k'}(\sigma,s)$, and thereby $\Lambda^{(l,l')}_{k,k'}$, by taking $\ket{\psi_1}\cong\ket{0,0}$ and $\ket{\psi_2}\cong\ket{1/2,-1/2}$. For the two eigenvalue energies, we used (\ref{eq:spin-frequency}) such that $\epsilon_1\cong\xi_{0,0}$ and $\epsilon_2\cong\xi_{1/2,-1/2}$. 

\begin{figure}[!htbp]
\center
\includegraphics[width=400pt, height=280pt]{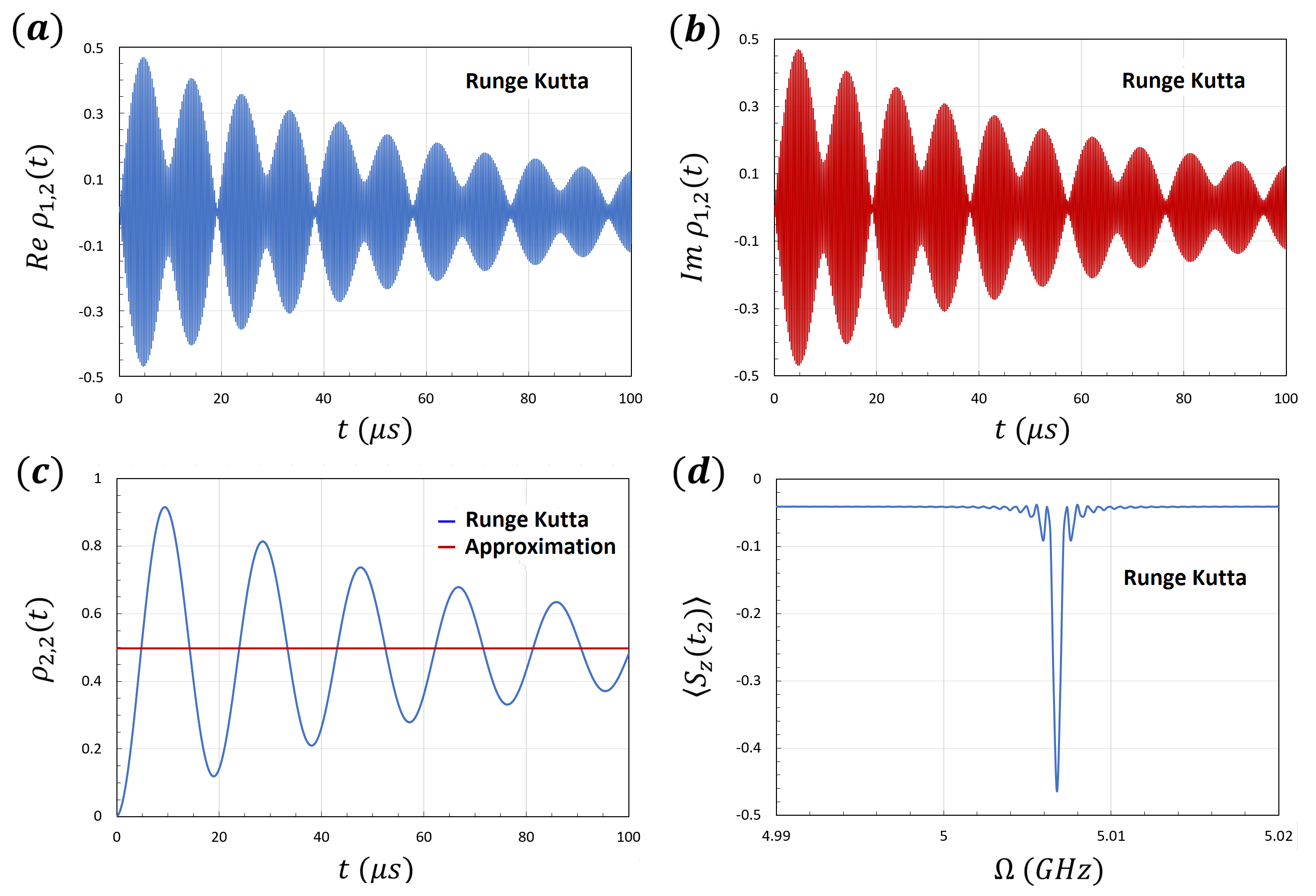}
\caption{\label{fig2} (a) Real and (b) imaginary parts of off-diagonal matrix element $\rho_{1,2}(t)$ plotted as a function of time $t$, where off-diagonal element $\rho_{2,1}(t)=\rho_{1,2}(t)^*$. (c) Diagonal matrix element $\rho_{2,2}(t)$ plotted as a function of time $t$, where $\rho_{1,1}(t)=1-\rho_{2,2}(t)$; the blue curve is the Runge Kutta numerical solution while the red is the asymptote of the steady-state limit, as $t\rightarrow\infty$, as discussed in the text. (d) Plot of Runge-Kutta estimated $\braket{S_z(t_2)}$, where $t_2=100$~$\mu$s, as a function of the frequency of the driving voltage, $\Omega$. Model parameters used to generate the results are described in the text.}
\end{figure}

Applying a Runge Kutta numerical method, we solved the four coupled differential equations to obtain the solution presented in Fig.~\ref{fig2}. With $\Omega\cong\epsilon_{2,1}$, corresponding to detuning of $50$~kHz, panels (a) and (b) of the figure show the real and imaginary parts of $\rho_{1,2}(t)$ as a function of $t$, respectively, while panel (c) displays the solution of $\rho_{2,2}(t)$ as a function of $t$. The other two elements of the density matrix are $\rho_{2,1}(t)=\rho_{1,2}(t)^*$ and $\rho_{1,1}(t)=1-\rho_{2,2}(t)$. Also, using (\ref{eq:qubit-expectation-value}), we plotted the $z$-component of the induced spin-1/2 qubit in panel (d), $\braket{S_z(t_2)}\cong -\rho_{2,2}(t)/2$, where $t_2=100~\mu$s, as a function of pump frequencies centered about the Rabi frequency; the $x$ and $y$ components of spin are zero and not displayed.

An approximate solution applicable to the steady-state, i.e., as $t\rightarrow\infty$, is derived in Section~V of the Supplemental Material \cite{Supplement2023}. In this approximation the off-diagonal density-matrix elements are purely oscillatory as a function of $t$, as $t\rightarrow\infty$, with an asymptotic form given by
\begin{equation}	\label{eq:rh-12}
\rho_{1,2}(t) \cong \frac{1}{2} \Bigg( \frac{\Lambda^{(2,2)}_{1,1} + \Lambda^{(1,1)}_{1,1}}{\Lambda^{(2,2)}_{1,1} - \Lambda^{(1,1)}_{1,1}} \Bigg) 
\Bigg[ \frac{ \theta_{1,2} \, \Lambda^{(1,2)}_{1,2} \, V_o \, e^{i\Omega t} }{ { ( \Omega - \tilde{\epsilon}_{2,1} ) }^2 + \Delta_{2,1}^2 } \Bigg] .
\end{equation}
In contrast, the leading contribution to the diagonal density-matrix elements are constant in $t$ as $t\rightarrow\infty$. For example, we have
\begin{equation}	\label{eq:rh-22}
\rho_{2,2}(t) \cong 
\frac{ -1 }{ \Lambda^{(2,2)}_{1,1} - \Lambda^{(1,1)}_{1,1} } \Bigg\{
\Lambda^{(1,1)}_{1,1}
+ \frac{1}{2} \Bigg( \frac{ \Lambda^{(2,2)}_{1,1} + \Lambda^{(1,1)}_{1,1} }{ \Lambda^{(2,2)}_{1,1} - \Lambda^{(1,1)}_{1,1} } \Bigg) 
\Bigg[ \frac{ {\left| \theta_{1,2} \right|}^2 \Lambda^{(1,2)}_{1,2} \, V_o^2 }{ { ( \Omega - \tilde{\epsilon}_{2,1} ) }^2 + \Delta_{2,1}^2 } \Bigg] \Bigg\} ,
\end{equation}
which is also plotted in Fig.~\ref{fig2}(c) as the red line. The asymptotes of $\rho_{1,1}(t)$ and $\rho_{2,2}(t)$ are the final, steady-state probability distributions achieved by virtue of the applied voltage. Also, within this approximation, $\braket{{\bf S}(t)}\cong-\rho_{2,2}(t)\hat{z}/2$, which implies
\begin{equation}	\label{eq:spin-z}
\braket{{\bf S}(t)}\cong
\frac{ 1 }{ 2 \left( \Lambda^{(2,2)}_{1,1} - \Lambda^{(1,1)}_{1,1} \right) } \Bigg\{
\Lambda^{(1,1)}_{1,1}
+ \frac{1}{2} \Bigg( \frac{ \Lambda^{(2,2)}_{1,1} + \Lambda^{(1,1)}_{1,1} }{ \Lambda^{(2,2)}_{1,1} - \Lambda^{(1,1)}_{1,1} } \Bigg) 
\Bigg[ \frac{ {\left| \theta_{1,2} \right|}^2 \Lambda^{(1,2)}_{1,2} \, V_o^2 }{ { ( \Omega - \tilde{\epsilon}_{2,1} ) }^2 + \Delta_{2,1}^2 } \Bigg] \Bigg\} \hat{z} .
\end{equation}

Note in the frequency response of (\ref{eq:rh-12}) through (\ref{eq:spin-z}) that the energy difference $\epsilon_{2,1}$ undergoes a small shift, by virtue of the applied voltage, to a new value
\begin{equation}	\label{eq:epsilon-21}
\tilde{\epsilon}_{2,1} = \epsilon_{2,1} 
- \left[ \frac{ 2 {\left| \theta_{1,2} \right|}^2 \Lambda^{(1,2)}_{1,2} + \Big( \theta_{1,2}^2 + \theta_{2,1}^2 \Big) \Lambda^{(2,1)}_{1,2} }
{ 4 \epsilon_{2,1} \Big( \Lambda^{(2,2)}_{1,1} - \Lambda^{(1,1)}_{1,1} \Big) } \right] V_o^2 .
\end{equation}
Also, the half-width at half maximum is
\begin{equation}	\label{eq:delta-21}
\Delta_{2,1} \cong \frac{1}{\tau} \sqrt{ 1 + \frac{1}{2} {\left| \theta_{1,2} \right|}^2 \, V_o^2 \, \tau^2 } \; ; \quad 
\tau = \frac{1}{\left| \Lambda^{(1,2)}_{1,2} \right|} ,
\end{equation}
where $\tau$ is a coherence time, with $\tau\cong 100.0~\mu$s in the present calculation. Equation (\ref{eq:delta-21}) is similar in form to that obtainable from the Bloch equation of motion \cite{Bloch1946}, except here $\tau$ is a single relaxation time governing the Lorentzian half-width, as opposed to the two longitudinal and transverse relaxation times of the Bloch equation. Dubois et al. \cite{Dubois_2021} have shown how the Bloch-like result follows from a specific semi-classical treatment of the Lindblad master equation. In the present quantum-limit result, (\ref{eq:rh-12}) through (\ref{eq:delta-21}) are most applicable when $t\gg\tau$.

\begin{figure}[!htbp]
\center
\includegraphics[width=400pt, height=250pt]{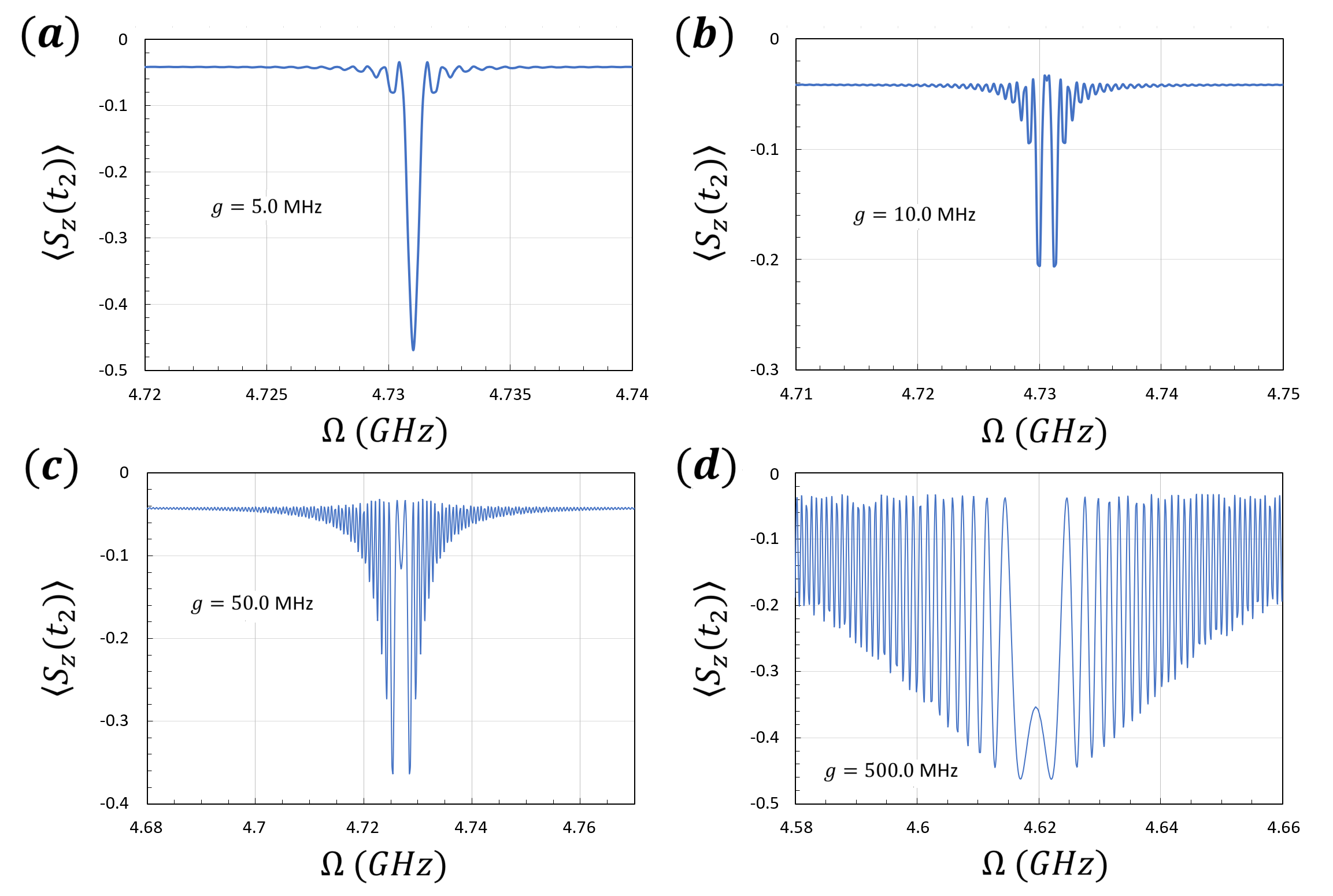}
\caption{\label{fig3} Plot of $\braket{S_z(t_2)}$, where $t_2=100$~$\mu$s, as a function of driving frequency $\Omega$, for several values of capacitive coupling parameter $g$, as indicated in the figure. Model parameters used to generate the results are described in the text.}
\end{figure}

As a last part of the analysis of the two-state excitation, we consider the Rabi resonance as a function of increasing $g$, using second-order perturbation theory to estimate $\ket{\psi_1}$ and $\ket{\psi_2}$, as well as $\epsilon_1$ and $\epsilon_2$. Specifically, in Fig.~\ref{fig3}, we plot the Runge-Kutta calculation of $\braket{S_z(t)}$ as a function of $\Omega$ for several increasing values of $g$, starting from $g=5$~GHz in panel (a). Note that Fig.~\ref{fig2}(d), of our previous calculation, and the new Fig.~\ref{fig3}(a) differ only in the estimate of $\ket{\psi_1}$, $\ket{\psi_2}$, $\epsilon_1$, and $\epsilon_2$. The greatest difference between the two plots is the resonance-frequency downward shift, by about $300$~MHz, in Fig.~\ref{fig3}(a), attributable to inclusion of the second-order perturbation.

With second-order perturbation theory applied in Fig.~\ref{fig3}, we are able to capture an avoided crossing involving the third and fourth energy levels, as discussed earlier with respect to Fig.~\ref{fig1}(c). Also, we assume the resonator interacts more strongly with the surrounding environment than does the transmon, with diffusion and dissipation rates given in Table~\ref{table1}. These factors allowed us to explore more fully the interplay between transmon-resonator states and polaritons that form in the presence of the driving voltage \cite{PhysRevA.93.063827,*PhysRevLett.120.083602}, particularly as we increase $g$.

In Fig.~\ref{fig3}, as we increase $g$ from $g=5.0$~GHz, in panel (a), to $g=10.0$~GHz, in panel (b), a gap initially opens in the frequency-response spectrum centered about $\Omega=\tilde{\epsilon}_{2,1}$, indicative of transparency at that frequency, with $\tilde{\epsilon}_{2,1}$ shifting lower as $g$ increases. At $g=10.0$~GHz, the transparency is nearly maximal, but as $g$ increases we see a non-monotonic character to the transparency, with an increasing number of sidebands appearing in the frequency response, which tend to overwhelm any apparent gap. For example, when $g=500$~MHz in Fig.~\ref{fig3}(d), the transparency has essentially disappeared. 

\subsection{A Two-Tone Three-State Rabi Resonance Calculation}
We next consider the solution of (\ref{eq:lindbladMasterEquation-subset}) for Rabi resonances between the three lowest-lying states, for which $\mathcal{E}=\left\{ 1, 2, 3 \right\}$. In this situation, matrix coefficients $\Theta_{k,k'}=\braket{\psi_k|\Theta\psi_{k'}}$, of the operator defined in (\ref{eq:voltage-coefficient}), couple the applied voltage to the two excited states, $\ket{\psi_2}$ and $\ket{\psi_3}$. However, since $\Theta_{1,3}=\Theta_{3,1}=0$, there can be no direct excitation of state $\ket{\psi_3}$ from the ground state. To overcome this forbidden transition, a ladder scheme can be employed to excite $\ket{\psi_1}\rightarrow\ket{\psi_2}$ and $\ket{\psi_2}\rightarrow\ket{\psi_3}$. This is accomplished by applying a two-tone voltage, $V(t)=V_p \, \sin{\Omega_p t}+V_c \, \sin{\Omega_c t}$. Here we set $V_p=0.5$~nV and $V_c=1.0$~nV, and adjust the two frequencies to obtain Rabi resonances $\Omega_p\approx\epsilon_{2,1}$ and $\Omega_c\approx\epsilon_{3,2}$, where $\epsilon_{k,k'}=\epsilon_k-\epsilon_{k'}$. For our calculations, we set the detuning to $50$~kHz for both Rabi resonances. The states $\ket{\psi_k}$ and $\epsilon_k$ were estimated from second-order perturbation theory, and a Runge-Kutta method was applied to solve the coupled differential equations of (\ref{eq:lindbladMasterEquation-subset}). 

\begin{figure}[!htbp]
\center
\includegraphics[width=400pt, height=250pt]{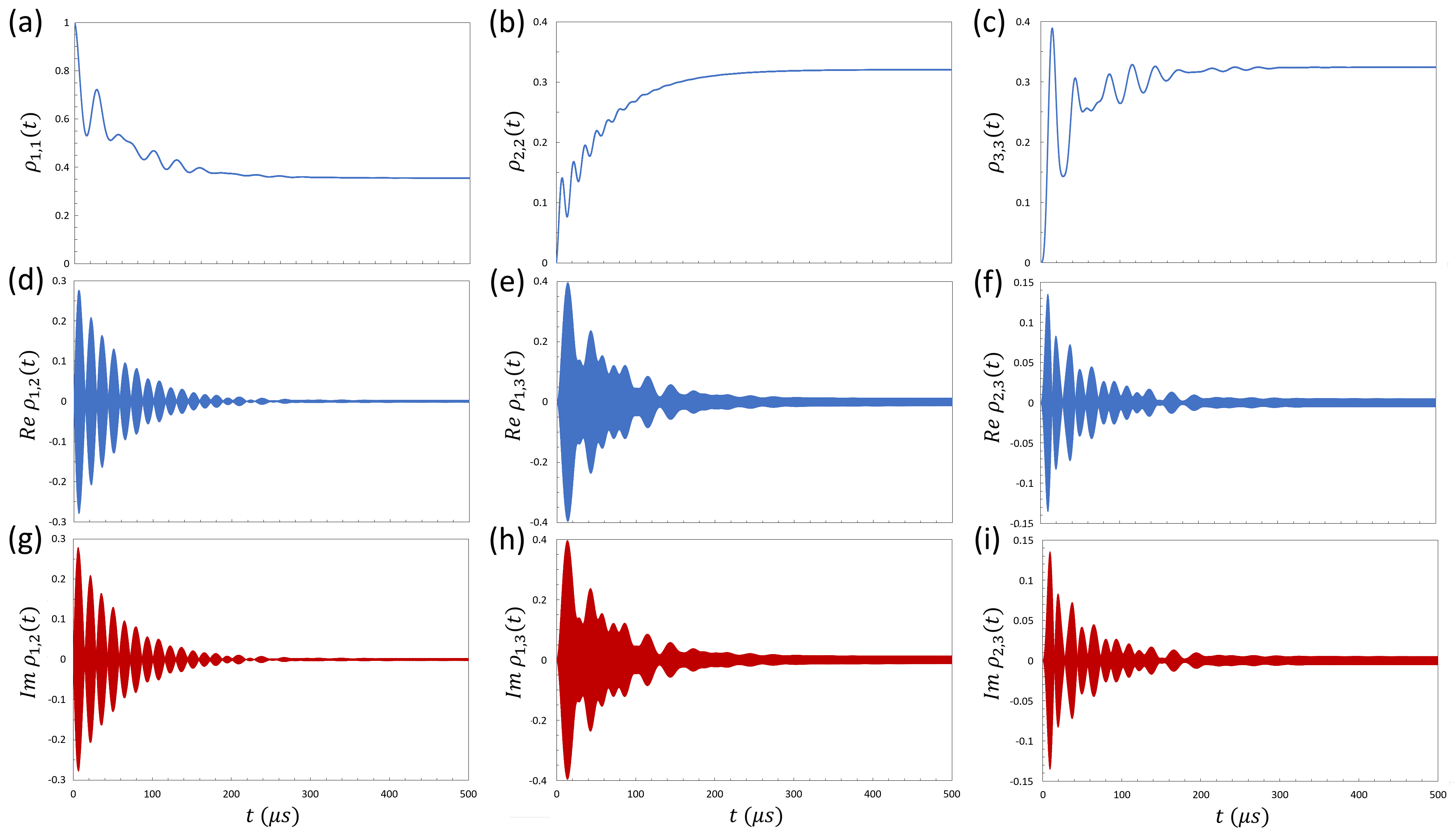}
\caption{\label{fig4} Plot of Runge-Kutta-evaluated elements of the density matrix $\rho(t)$ of the ladder-type state-excitation scheme as a function of time $t$, as discussed in the text. Panels (a) through (c) represent the diagonal elements, panels (d) through (f) depict the real parts of off-diagonal elements $\rho_{1,2}(t)$, $\rho_{1,3}(t)$, and $\rho_{2,3}(t)$, respectively. The imaginary parts of these same matrix elements are shown plotted in red in panels (g) through (i). Model parameters used to generate the results are described in the text.}
\end{figure}

Figure~\ref{fig4} displays the solution of the density matrix elements as a function of time for the ladder-type excitation scheme. Panels (a) through (c) display the behavior of the diagonal elements and the approach to a steady-state distribution of qubit spin. The final, steady-state wave function of the system is comprised of a constant admixture of states $S=0$, $S=1/2$, and $S=1$, of approximately $1/3$ probability for each. The real parts of $\rho_{1,2}(t)$, $\rho_{1,3}(t)$, and $\rho_{2,3}(t)$ are shown as a function of time in panels (d) through (f), respectively. Similarly, imaginary parts of $\rho_{1,2}(t)$, $\rho_{1,3}(t)$, and $\rho_{2,3}(t)$ are shown as a function of time in panels (g) through (i), respectively. Other off-diagonal density-matrix elements are $\rho_{2,1}(t)=\rho_{1,2}(t)^*$, $\rho_{3,1}(t)=\rho_{1,3}(t)^*$, and $\rho_{3,2}(t)=\rho_{2,3}(t)^*$. 

\section{Conclusion\label{sec:conclusion}}
In this study we have proposed a way to define the qubit of a superconducting transmon by leveraging its proximity to a resonator readout. In this approach the qubit is defined as a measure of photon entanglement, via the use of the Schwinger oscillator model of angular momentum. The Schwinger oscillator construct of angular momentum provides a natural and robust description of a qubit, applicable to a multilevel model of transmon and resonator energy states. It is not restricted to a spin-1/2 representation, and can be considered a definition for an emergent property of the combined transmon-resonator system. 

The Schwinger construct introduces angular-momentum, atom-like symmetry, corresponding to underlying conservation of total photon number, $N$, as a basis from which to construct the combined transmon-resonator wave function, and thereby, the expectation value of the qubit. This basis provides a convenient starting point from which to study the error-inducing effects of transmon anharmonicity, surrounding-environment decoherence, and random stray fields on qubit state. Specifically, in this analysis, we showed that, even in the presence of weak transmon anharmonicity, selection rules exist for photon state transitions. In particular, we showed, via Fig.~\ref{fig1}(b), that transitions between states that change $S=N/2$ by a half-integer, or equivalently, change $N$ by an odd number, are forbidden. This implies, for example, that the superconducting transmon tends to be robust against random one-half noise events because such events represent forbidden one-photon transitions. As transom anharmonicity increases, and the effects of symmetry breaking become stronger, these forbidden transitions become less strict.

\begin{acknowledgments}
We thank D. Pappas, V. Markov, and C. Gonciulea for helpful comments and suggestions, and additional support and encouragement from S. Holston and T. Skeen. The views expressed in this article are those of the authors and do not represent the views of Wells Fargo. This article is for informational purposes only. Nothing contained in this article should be construed as investment advice. Wells Fargo makes no express or implied warranties and expressly disclaims all legal, tax, and accounting implications related to this article.
\end{acknowledgments}

\bibliography{bibliography}

\ifarXiv
    

\section*{Supplementary material}

# Supplement to Application of the Schwinger Oscillator Model of Angular Momentum to an Interpretation of the Superconducting Transmon Qubit

R. P. Erickson[1, *]

[1]*Wells Fargo*

(Dated: January 17, 2024)

## Abstract

Supplemental details are provided regarding a study of the Schwinger oscillator model of angular momentum, applied to the superconducting transmon and its resonator readout. In these notes, we discuss the transmon/resonator Hamiltonian, the Schwinger oscillator model of angular momentum applied to a definition of the qubit, and the two-state Rabi resonance of this qubit in the steady state.

## I. INTRODUCTION

This supplement presents several derivations associated with the application of the Schwinger oscillator model of angular momentum [1] to the definition of a superconducting transmon qubit. The transmon is coupled to a resonator and driven by an external voltage. Specifically, we discuss the transmon/resonator Hamiltonian, the Schwinger oscillator model of angular momentum within the scope of the qubit definition, and an approximation of a two-state Rabi resonance of this qubit in the steady state. In the Rabi resonance approximation, we calculate the frequency response and coherence time of the qubit using a Lindblad master equation, which models the immersion of both the transmon and resonator within the surrounding environment.

## II. THE MODEL OF THE TRANSMON AND RESONATOR READOUT

In this study we consider a superconducting transmon of charging energy $E_C$, Josephson energy $E_J$, and flux quantum, $\Phi_o = \hbar/2e$, that can be modeled as an anharmoinc oscillator. The periodic energy of flux is given by $-E_J \cos\varphi_-$, where we have defined $\varphi_- = \phi_J/\Phi_o$, with $\phi_J$ a Josephson phase. We count the quanta of charge, $q_-$, by $n_- = -q_-/2e$, which is the number of Cooper pairs that cross the junction. Accounting for a charge offset, via number $n_g$, with origin in a capacitively coupled gate electrode and/or other influences, we have a Hamiltonian that can be expressed as

$$\mathscr{H}_- = 4E_C(n_- - n_g)^2 - E_J \cos\varphi_-. \tag{1}$$

---
* Electronic address: Robert.Erickson2@wellsfargo.com

For the superconducting transmon, with weak anharmonicity, we can approximate the flux via a Taylor expansion of the cosine to fourth-order in $\varphi_-$. Also, assuming the potential well is deep about $\varphi_- = 0$, so photons of lowest-lying states are most probable there, we can eliminate the $n_g$ term of (1) via a suitable transformation [2]. The end result is a Duffing approximation given by

$$\mathcal{H}_- \cong 4E_C\, n_-^2 + \frac{1}{2} E_J\, \varphi_-^2 - \frac{1}{24} E_J\, \varphi_-^4, \tag{2}$$

where we have shifted the Hamiltonian to remove constant terms that have no bearing on dynamics.

We also consider a transmission line for readout, modeled as a linear resonator, where the self inductance and capacitance are defined as $L$ and $C$, respectively. Here, the equivalent $LC$-circuit charge and phase are denoted by $q_+$ and $\varphi_+$, respectively, such that the Hamiltonian is

$$\mathcal{H}_+ = \frac{1}{2L}\varphi_+^2 + \frac{1}{2C} q_+^2. \tag{3}$$

We assume the transmission line is capacitively coupled to the transmon, which introduces an interaction term of the form

$$\mathcal{H}_{+,-} = \beta\, q_+\, n_-, \tag{4}$$

where $\beta$ is a proportionality constant. If the full Hamiltonian is defined as $\mathcal{H}_o = \mathcal{H}_+ + \mathcal{H}_- + \mathcal{H}_{+,-}$ then we have

$$\mathcal{H}_o = \frac{1}{2L}\varphi_+^2 + \frac{1}{2C} q_+^2 + 4E_C\, n_-^2 + \frac{1}{2} E_J\, \varphi_-^2 - \frac{1}{24} E_J\, \varphi_-^4 + \beta\, q_+\, n_-. \tag{5}$$

As a last step, we allow the system of $\mathcal{H}_o$ to be driven by an external voltage, introducing a time-dependent term, acting on the transmission line, that takes the form

$$\mathcal{U}(t) = -q_+\, V(t), \tag{6}$$

where $V(t)$ is an effective driving voltage. Thus, the complete open-system Hamiltonian may be expressed as

$$\mathcal{H}(t) = \mathcal{H}_o + \mathcal{U}(t). \tag{7}$$

The Hamiltonian of (7) is quantized by allowing conjugate pairs $\varphi_+$, $q_+$ and $n_-$, $\varphi_-$ to denote quantum operators, where $[q_+, \varphi_+] = i\hbar$ and $[\varphi_-, n_-] = i$. These can be further second-quantized by defining

$$q_+ = \sqrt{\frac{\hbar}{2Z}}\left(a_+^\dagger + a_+\right), \quad \phi_+ = i\sqrt{\frac{\hbar Z}{2}}\left(a_+^\dagger - a_+\right), \tag{8}$$

$$n_- = i\sqrt{\frac{\alpha}{2}}\left(a_-^\dagger - a_-\right), \quad \varphi_- = \frac{1}{\sqrt{2\alpha}}\left(a_-^\dagger + a_-\right), \quad (9)$$

where

$$Z = \sqrt{\frac{L}{C}}, \quad \alpha = \sqrt{\frac{E_J}{8E_C}}. \quad (10)$$

Here, the conjugate variables are expressed in terms of pairs of annihilation and creation operators $a_+$, $a_+^\dagger$ and $a_-$, $a_-^\dagger$. Introducing indexes $\mu, \mu' = \pm$, these new ladder operators have boson commutation relations $[a_\mu, a_{\mu'}^\dagger] = \delta_{\mu,\mu'}$, $[a_\mu, a_{\mu'}] = 0$, and $[a_\mu^\dagger, a_{\mu'}^\dagger] = 0$. Substituting (8) and (9) into (5) and (6), we arrive at

$$\mathcal{H}_o = \sum_{\mu=\pm} \hbar\omega_\mu \left(a_\mu^\dagger a_\mu + \frac{1}{2}\right) + \frac{1}{4}i\hbar\tilde{g}\left(a_+^\dagger + a_+\right)\left(a_-^\dagger - a_-\right) - \frac{1}{12}E_C\left(a_-^\dagger + a_-\right)^4, \quad (11)$$

$$\mathcal{U}(t) = -\sqrt{\frac{\hbar}{2Z}}\left(a_+^\dagger + a_+\right)V(t), \quad (12)$$

where

$$\omega_+ = \frac{1}{\sqrt{LC}}, \quad \omega_- = \frac{1}{\hbar}\sqrt{8E_C E_J}. \quad (13)$$

Also, in the coupling between oscillators, we have introduced coefficients

$$g = \frac{\beta}{\sqrt{2\hbar Z}}, \quad \tilde{g} = g\sqrt{\frac{\hbar\omega_-}{E_C}}. \quad (14)$$

### III. THE CANONICAL FORM OF THE HAMILTONIAN

As shown in Appendix A, it is useful to affect a canonical transformation on the quadratic terms of $\mathcal{H}_o$, rendering them diagonal in a new space of decoupled oscillators, denoted by $\sigma = \uparrow$ and $\sigma = \downarrow$. Via (A6), the fundamental frequencies of these canonical oscillators can be expressed succinctly as

$$\omega_\sigma = \sqrt{\frac{1}{2}\left[\omega_+^2 + \omega_-^2 + \sigma\sqrt{(\omega_+^2 - \omega_-^2)^2 + \tilde{g}^2\,\omega_+\omega_-}\right]}, \quad (15)$$

where $\sigma = \uparrow$ ($\sigma = \downarrow$) has the algebraic meaning of integer $+1$ ($-1$) within the square brackets.

In this transformed space, new annihilation and creation operators $a_\sigma$ and $a_\sigma^\dagger$ act on the energy levels of the canonical oscillators and satisfy commutation relations like those of $a_\mu$ and $a_\mu^\dagger$. Via (A11), the transformation and its inverse can be expressed, respectively, as

$$A_{\sigma,s} = \sum_{\mu,m} u_{\sigma,s}(\mu, m)\, A_{\mu,m}, \quad A_{\mu,m} = \sum_{\sigma,s} e_{\sigma,s}(\mu, m)\, A_{\sigma,s}, \quad (16)$$

where indexes $m = \pm$ and $s = \pm$. Here we have adopted a notation where the ladder operators are expressed as jump operators $A_{\mu,+} = a_\mu$, $A_{\mu,-} = a_\mu^\dagger$, $A_{\sigma,+} = a_\sigma$, and $A_{\sigma,-} = a_\sigma^\dagger$.

Via (A7), the coefficients of the transformation, elements of right and left eigenvectors, respectively, are

$$e_{\sigma,s}(\mu, m) = \frac{1}{2}\xi_{\sigma,\mu}\left(\frac{\omega_\mu + ms\,\omega_\sigma}{\omega_\sigma}\right) e^{i\pi[\sigma m(1-\mu)-s(1-\sigma)]/4}, \quad u_{\sigma,s}(\mu, m) = ms\, e_{\sigma,s}(\mu, m)^*, \quad (17)$$

where $\mu$, $m$, and $s$ appear algebraically as integers $\pm 1$. The coefficient $\xi_{\sigma,\mu}$ is given in (A8) as

$$\xi_{\sigma,\mu} = \sqrt{\frac{\omega_\sigma}{\omega_\mu}\left(\frac{\omega_{\bar{\mu}}^2 - \omega_\sigma^2}{\omega_{\bar{\sigma}}^2 - \omega_\sigma^2}\right)}, \quad (18)$$

where $\bar{\mu} = -\mu$ and $\bar{\sigma} = \downarrow$ ($\bar{\sigma} = \uparrow$) when $\sigma = \uparrow$ ($\sigma = \downarrow$). When (16) is applied to (11) we obtain the canonical Hamiltonian [3], which we can express as

$$\mathcal{H}_o = \sum_{\sigma \in \{\uparrow,\downarrow\}} \hbar\omega_\sigma \left(a_\sigma^\dagger a_\sigma + \frac{1}{2}\right) - \frac{1}{12}E_c\left[\xi_{\uparrow,-}\left(a_\uparrow^\dagger - a_\uparrow\right) - i\xi_{\downarrow,-}\left(a_\downarrow^\dagger + a_\downarrow\right)\right]^4. \quad (19)$$

Similarly, the canonical form of the driving term of (12) can be expressed as

$$\mathcal{U}(t) = -\hbar\Theta\, V(t), \quad (20)$$

$$\Theta = v_\uparrow \left(a_\uparrow^\dagger + a_\uparrow\right) + iv_\downarrow \left(a_\downarrow^\dagger - a_\downarrow\right); \quad v_\sigma = \frac{1}{\sqrt{2\hbar Z}}\left(\frac{\omega_+}{\omega_\sigma}\right)\xi_{\sigma,+}. \quad (21)$$

## IV. THE SCHWINGER OSCILLATOR MODEL OF ANGULAR MOMENTUM

With canonical transformation applied to the Hamiltonian, we now introduce a metric to describe entanglement of photons within the combined transmon and resonator system. Specifically, following the method of Schwinger [1], we use the one-particle canonical ladder operators to define corresponding two-particle angular momentum operators

$$S_+ = a_\uparrow^\dagger a_\downarrow, \quad S_- = a_\downarrow^\dagger a_\uparrow, \quad (22)$$

$$S_x = \frac{1}{2}(S_+ + S_-), \quad S_y = \frac{1}{2i}(S_+ - S_-), \quad S_z = \frac{1}{2}\left(a_\uparrow^\dagger a_\uparrow - a_\downarrow^\dagger a_\downarrow\right), \quad (23)$$

which describe a pseudo spin $\mathbf{S} = S_x\hat{x} + S_y\hat{y} + S_z\hat{z}$ of length $S$, where $\hat{x}$, $\hat{y}$, and $\hat{z}$ are unit vectors of an abstract Cartesian coordinate system. The pseudo spin $\mathbf{S}$ is defined dimensionless since it is not a physical representation of angular momentum. The quantum number $S$ is related to the total photon number $N$ by the expressions

$$S = N/2, \quad N = n_\uparrow + n_\downarrow, \quad n_\sigma = a_\sigma^\dagger a_\sigma. \quad (24)$$

An excellent description of the Schwinger oscillator model of angular momentum can be found in the text of J. J. Sakurai [4].

The spin components $S_x$, $S_y$, and $S_z$ satisfy the usual commutation relations of angular momentum, i.e., $[S_j, S_k] = i\epsilon_{j,k,l} S_l$, where $j, k, l \in \{x, y, z\}$ and $\epsilon_{j,k,l}$ is a Levi-Civita coefficient, via the commutation relations of the canonical ladder operators. While we have introduced the pseudo-spin components via $a_\sigma$ and $a_\sigma^\dagger$, we can also write these components in terms of the original $a_\mu$ and $a_\mu^\dagger$ by substituting the inverse canonical transformation of (16) into (22) and (23). The Schwinger spin is well defined in either original or canonical space since the spin commutation relations are preserved.

In the linear limit of (19), energy eigenstates are two-particle states of the form $|n_\uparrow\rangle \otimes |n_\downarrow\rangle$, where $a_\sigma |n_\sigma\rangle = \sqrt{n_\sigma} |n_\sigma - 1\rangle$ and $a_\sigma^\dagger |n_\sigma\rangle = \sqrt{n_\sigma + 1} |n_\sigma + 1\rangle$, with eigenvalues $E_{n_\uparrow, n_\downarrow} = \sum_\sigma \hbar \omega_\sigma (n_\sigma + 1/2)$. In this limit, angular momentum states are identical to two-particle energy eigenstates, but with reordered indexing, such that an angular momentum state is of the form

$$|S, m_S\rangle = \frac{a_\uparrow^{\dagger\, S+m_S} |0\rangle \otimes a_\downarrow^{\dagger\, S-m_S} |0\rangle}{\sqrt{(S+m_S)!(S-m_S)!}}, \quad (25)$$

where $|0, 0\rangle = |0\rangle \otimes |0\rangle$ is the vacuum state, length $S = (n_\uparrow + n_\downarrow)/2 = 0, 1/2, 1, \ldots$, and quantum number $m_S = (n_\uparrow - n_\downarrow)/2 = -S, -S+1, \ldots, S-1, S$. The state $|S, m_S\rangle$ is an eigenstate of the linear Hamiltonian, with eigenvalue $\hbar \xi_{S, m_S}$, where

$$\xi_{S, m_S} = (\omega_\uparrow + \omega_\downarrow)\left(S + \frac{1}{2}\right) + (\omega_\uparrow - \omega_\downarrow) m_S. \quad (26)$$

Also, since operators $S_\pm$ of (22) are legitimate angular momentum operators, we have the identities

$$S_\pm |S, m_S\rangle = \sqrt{(S \mp m_S)(S \pm m_S + 1)} |S, m_S \pm 1\rangle, \quad (27)$$

as the reader can verify.

The states $|S, m_S\rangle$ of (25) are symmetric since the particles of the combined oscillators are bosons. It is useful to express the Hamiltonian of (19) in this representation, i.e., given the eigenproblem

$$\mathcal{H}_o |\psi_k\rangle = \hbar \epsilon_k |\psi_k\rangle, \quad (28)$$

we can write

$$\sum_{S', m_S'} \langle S, m_S | \mathcal{H}_o | S', m_S' \rangle \langle S', m_S' | \psi_k \rangle = \hbar \epsilon_k \langle S, m_S | \psi_k \rangle \; ; \quad k = 0, 1, 2, \ldots \quad (29)$$

In the case $E_C \ll \omega_\uparrow, \omega_\downarrow$ in (19), the set of energies $\hbar \epsilon_k$ is only slightly perturbed from that of $\hbar \xi_{S,m_S}$, indicative of minor angular-momentum symmetry breaking, allowing $\epsilon_k$ and $\langle S, m_S | \psi_k \rangle$ to be estimated using standard Rayleigh-Schrödinger time-independent perturbation theory [5].

## V. THE SCHWINGER QUBIT NEAR A RABI RESONANCE

We model the immersion of the transmon and resonator in the surrounding environment via a Lindblad master equation [6], where $\gamma'_\mu$ and $\gamma_\mu$ are diffusion and dissipation rates, respectively, between environment and resonator ($\mu = +$), and environment and transmon ($\mu = -$). In this model we assume $|\tilde{g}| \ll 4\omega_\mu$ such that, to first approximation, we can neglect diffusion and dissipation directly between transmon and resonator. Under the canonical transformation of (16) and (17), the Lindblad master equation can be expressed as

$$\frac{d\rho(t)}{dt} = \frac{1}{i\hbar} [\mathscr{H}(t), \rho(t)] + \frac{1}{2} \sum_{\sigma,\sigma' \in \{\uparrow,\downarrow\}} \sum_{s,s'=\pm} \gamma_{\sigma,s;\sigma',s'} \left[ 2 A_{\sigma,s}\, \rho(t)\, A^\dagger_{\sigma',s'} - A^\dagger_{\sigma',s'} A_{\sigma,s}\, \rho(t) - \rho(t)\, A^\dagger_{\sigma',s'} A_{\sigma,s} \right], \quad (30)$$

where $\rho(t)$ is a density operator, $A_{\sigma,s}$ is a canonical jump operator, as in (16), and

$$\gamma_{\sigma,s;\sigma',s'} = \frac{1}{4} \sum_{\mu,m=\pm} \gamma_{\mu,m}\, \xi_{\sigma,\mu}\, \xi_{\sigma',\mu} \left( \frac{\omega_\mu + m s\, \omega_\sigma}{\omega_\sigma} \right) \left( \frac{\omega_\mu + m s'\, \omega_{\sigma'}}{\omega_{\sigma'}} \right)$$
$$\times e^{i\pi[(\sigma-\sigma')m(1-\mu)-s(1-\sigma)+s'(1-\sigma')]/4}, \quad (31)$$

with $\gamma_{\mu,m} = (\gamma'_\mu + \gamma_\mu)\, \delta_{m=+} + (\gamma'_\mu - \gamma_\mu)\, \delta_{m=-}$.

To solve (30), we write it in component form as a set of coupled differential equations in matrix elements $\rho_{k,k'}(t) = \langle \psi_k | \rho(t) | \psi_{k'} \rangle$, where we also define $\Theta_{k,k'} = \langle \psi_k | \Theta | \psi_{k'} \rangle$, for $\Theta$ of (21), and $A_{k,k'}(\sigma, s) = \langle \psi_k | A_{\sigma,s} | \psi_{k'} \rangle$, for jump operator $A_{\sigma,s}$. We assume a subset of states, denoted by $\mathscr{E}$, couple strongly to the applied voltage $V(t)$ near a Rabi resonance, such that, to good approximation, we effectively can set $\rho_{k,k'}(t) = 0$ unless $k, k' \in \mathscr{E}$. In this way, we arrive at a practical expression of the Lindblad master equation given by

$$\frac{d\rho_{k,k'}(t)}{dt} \cong -i\,(\epsilon_k - \epsilon_{k'})\, \rho_{k,k'}(t) + \sum_{l,l' \in \mathscr{E}} \left[ i\, (\Theta_{k,l}\, \delta_{k',l'} - \Theta_{l',k'}\, \delta_{k,l})\, V(t) + \Lambda^{(l,l')}_{k,k'} \right] \rho_{l,l'}(t), \quad (32)$$

where we have defined canonical coefficients

$$\Lambda_{k,k'}^{(l,l')} = \frac{1}{2} \sum_{\sigma,\sigma' \in \{\uparrow,\downarrow\}} \sum_{s,s'=\pm} \gamma_{\sigma,s;\sigma',s'} \Bigg\{ 2 A_{k,l}(\sigma,s) \, A_{k',l'}(\sigma',s')^*$$
$$- \sum_{k'' \in \mathscr{E}} [A_{k'',k}(\sigma',s')^* \, A_{k'',l}(\sigma,s) \, \delta_{k',l'} + A_{k'',l'}(\sigma',s')^* \, A_{k'',k'}(\sigma,s) \, \delta_{k,l}] \Bigg\}, \quad (33)$$

indicative of the strength of interaction with the surroundings.

Given Schwinger spin component $S_{k,k'}^{(\alpha)} = \langle \psi_k | S_\alpha | \psi_{k'} \rangle$, where $\alpha \in \{x, y, z\}$, $\hat{\alpha}$ is a Cartesian unit vector, and the expectation value of the spin is $\langle \mathbf{S}(t) \rangle = \text{Tr}\,[\mathbf{S}\rho(t)]$, we can write

$$\langle \mathbf{S}(t) \rangle \cong \sum_{\alpha \in \{x,y,z\}} \sum_{k,k' \in \mathscr{E}} S_{k,k'}^{(\alpha)} \, \rho_{k',k}(t) \, \hat{\alpha}, \quad (34)$$

with $\rho_{k',k}(t)$ determined from (32). In what follows, we consider a single-tone voltage $V(t) = V_o \sin \Omega t$, of frequency $\Omega$ and amplitude $V_o$, that excites the transmon-resonator system into a Rabi resonance, between the ground state and the lowest-lying spin-1/2 state. We derive an approximate solution for $\rho_{k',k}(t)$ in the steady state, i.e., as $t \to \infty$, and we apply this solution to an estimate of $\langle \mathbf{S}(t) \rangle$, via (34).

### A. The Approximate Two-State Density Matrix

We show how to obtain an approximate solution of (32), where $\mathscr{E} = \{1, 2\}$, as $t \to \infty$. The two states $|\psi_1\rangle$ and $|\psi_2\rangle$ correspond to energy eigenstates $\hbar\epsilon_1$ and $\hbar\epsilon_2$, respectively, where we define $\epsilon_{2,1} = \epsilon_2 - \epsilon_1 > 0$. We assume a near-resonance condition such that $\Omega \approx \epsilon_{2,1}$. The coupled equations of (32) are then

$$\frac{d}{dt}\rho_{1,2}(t) \cong -i\theta_{1,2}\, V(t) + \left(i\epsilon_{2,1} + \Lambda_{1,2}^{(1,2)}\right)\rho_{1,2}(t) + \Lambda_{1,2}^{(2,1)}\, \rho_{2,1}(t) + 2i\,\theta_{1,2}\, V(t)\, \rho_{2,2}(t), \quad (35)$$

$$\frac{d}{dt}\rho_{2,1}(t) \cong i\theta_{2,1}\, V(t) + \left(-i\epsilon_{2,1} + \Lambda_{1,2}^{(1,2)}\right)\rho_{2,1}(t) + \Lambda_{1,2}^{(2,1)}\, \rho_{1,2}(t) - 2i\,\theta_{2,1}\, V(t)\, \rho_{2,2}(t), \quad (36)$$

$$\frac{d}{dt}\rho_{2,2}(t) \cong -\Lambda_{1,1}^{(1,1)} - \left(\Lambda_{1,1}^{(2,2)} - \Lambda_{1,1}^{(1,1)}\right)\rho_{2,2}(t) + i\theta_{2,1}\, V(t)\, \rho_{1,2}(t) - i\theta_{1,2}\, V(t)\, \rho_{2,1}(t), \quad (37)$$

where we note that $\rho_{1,1}(t) = 1 - \rho_{2,2}(t)$ via conservation of probability. In this approximation we estimate coefficients $\Theta_{k,k'}$ and $\Lambda_{k,k'}^{(l,l')}$ using $|\psi_1\rangle \cong |0, 0\rangle$ and $|\psi_2\rangle \cong |1/2, -1/2\rangle$, with $\epsilon_1 \cong \xi_{0,0}$ and $\epsilon_2 \cong \xi_{1/2,-1/2}$.

A rotating-wave approximation [7] can be applied to the above coupled equations, but a more general approach, applicable when there are multiple Rabi resonances, is used here,

one that assumes a solution in the form of a Fourier-like expansion in harmonics of $\Omega$ [8]. As $t \to \infty$, corresponding to the steady state, the Fourier-like coefficients can be considered time independent, such that the solution takes the form

$$\rho_{1,2}(t) = \sum_n \rho_{1,2}^{(n)} e^{in\Omega t}, \quad \rho_{2,1}(t) = \sum_n \rho_{2,1}^{(n)} e^{in\Omega t}, \quad \rho_{2,2}(t) = \sum_n \rho_{2,2}^{(n)} e^{in\Omega t}. \quad (38)$$

Note that our solution will not incorporate the boundary condition at $t = 0$ since (38) is only applicable for $t \gg 0$. Substituting (38) into (35) through (37), and equating terms of the same harmonic in $\Omega$, we arrive at

$$\left[i\left(\Omega n - \epsilon_{2,1}\right) - \Lambda_{1,2}^{(1,2)}\right]\rho_{1,2}^{(n)} - \Lambda_{1,2}^{(2,1)}\rho_{2,1}^{(n)} + \theta_{1,2}V_o\left(\rho_{2,2}^{(n+1)} - \rho_{2,2}^{(n-1)}\right) \cong -\frac{1}{2}\theta_{1,2}V_o\left(\delta_{n,1} - \delta_{n,-1}\right), \quad (39)$$

$$\left[i\left(\Omega n + \epsilon_{2,1}\right) - \Lambda_{1,2}^{(1,2)}\right]\rho_{2,1}^{(n)} - \Lambda_{1,2}^{(2,1)}\rho_{1,2}^{(n)} - \theta_{2,1}V_o\left(\rho_{2,2}^{(n+1)} - \rho_{2,2}^{(n-1)}\right) \cong \frac{1}{2}\theta_{2,1}V_o\left(\delta_{n,1} - \delta_{n,-1}\right), \quad (40)$$

$$\left(i\Omega n + \Lambda_{1,1}^{(2,2)} - \Lambda_{1,1}^{(1,1)}\right)\rho_{2,2}^{(n)} + \frac{1}{2}\theta_{2,1}V_o\left(\rho_{1,2}^{(n+1)} - \rho_{1,2}^{(n-1)}\right) - \frac{1}{2}\theta_{1,2}V_o\left(\rho_{2,1}^{(n+1)} - \rho_{2,1}^{(n-1)}\right) \cong -\Lambda_{1,1}^{(1,1)}\delta_{n,0}. \quad (41)$$

Again, these expressions only apply for $t \gg 0$; they determine the final distribution of states created by the Rabi resonance, which is are primary interest.

Since $\Omega \approx \epsilon_{2,1}$, resonance is with the principal harmonic of the applied voltage, so we expect coefficients of $n = \pm 1$ to be most significant in the above equations, with coefficients of $|n| > 1$ small enough in magnitude to be neglected in the first approximation. For example, using this assumption we can rearrange (41) to obtain

$$\rho_{2,2}^{(0)} \cong \frac{-1}{\Lambda_{1,1}^{(2,2)} - \Lambda_{1,1}^{(1,1)}}\left[\Lambda_{1,1}^{(1,1)} + \frac{1}{2}\theta_{2,1}V_o\left(\rho_{1,2}^{(1)} - \rho_{1,2}^{(-1)}\right) - \frac{1}{2}\theta_{1,2}V_o\left(\rho_{2,1}^{(1)} - \rho_{2,1}^{(-1)}\right)\right], \quad (42)$$

$$\rho_{2,2}^{(-1)} \cong -\frac{1}{2}V_o\left(\frac{\theta_{2,1}\rho_{1,2}^{(0)} - \theta_{1,2}\rho_{2,1}^{(0)}}{-i\Omega + \Lambda_{1,1}^{(2,2)} - \Lambda_{1,1}^{(1,1)}}\right), \quad \rho_{2,2}^{(+1)} \cong \frac{1}{2}V_o\left(\frac{\theta_{2,1}\rho_{1,2}^{(0)} - \theta_{1,2}\rho_{2,1}^{(0)}}{i\Omega + \Lambda_{1,1}^{(2,2)} - \Lambda_{1,1}^{(1,1)}}\right). \quad (43)$$

Thus, we have an approximate solution of the relevant diagonal density matrix elements given a solution of their off-diagonal counterparts. Also, note that these equations imply

$$\rho_{2,2}^{(1)} - \rho_{2,2}^{(-1)} \cong \frac{V_o\left(\Lambda_{1,1}^{(2,2)} - \Lambda_{1,1}^{(1,1)}\right)\left(\theta_{2,1}\rho_{1,2}^{(0)} - \theta_{1,2}\rho_{2,1}^{(0)}\right)}{\Omega^2 + \left(\Lambda_{1,1}^{(2,2)} - \Lambda_{1,1}^{(1,1)}\right)^2}. \quad (44)$$

For the case $n = 0$, if we substitute (44) into (39) and (40) we find two homogeneous coupled algebraic equations in coefficients $\rho_{1,2}^{(0)}$ and $\rho_{2,1}^{(0)}$, which requires us to conclude that $\rho_{1,2}^{(0)} = \rho_{2,1}^{(0)} = 0$. Thus, via (43), we also have $\rho_{2,2}^{(\pm 1)} = 0$.

9

We are left to solve for the remaining coefficients, $\rho_{1,2}^{(\pm 1)}$ and $\rho_{1,2}^{(\pm 1)}$. To obtain these solutions, we set $n = \pm 1$ in (39) and (40), and replace $\rho_{2,2}^{(0)}$ with the expression on the right side of (42). This gives the preliminary set of four coupled algebraic equations

$$\left[i\left(\Omega - \epsilon_{2,1}\right) - \Lambda_{1,2}^{(1,2)} + \frac{|\theta_{1,2}|^2 V_o^2}{2\left(\Lambda_{1,1}^{(2,2)} - \Lambda_{1,1}^{(1,1)}\right)}\right]\rho_{1,2}^{(1)} - \left[\Lambda_{1,2}^{(2,1)} + \frac{\theta_{1,2}^2 V_o^2}{2\left(\Lambda_{1,1}^{(2,2)} - \Lambda_{1,1}^{(1,1)}\right)}\right]\rho_{2,1}^{(1)}$$
$$- \frac{|\theta_{1,2}|^2 V_o^2}{2\left(\Lambda_{1,1}^{(2,2)} - \Lambda_{1,1}^{(1,1)}\right)}\rho_{1,2}^{(-1)} + \frac{\theta_{1,2}^2 V_o^2}{2\left(\Lambda_{1,1}^{(2,2)} - \Lambda_{1,1}^{(1,1)}\right)}\rho_{2,1}^{(-1)} \cong -\frac{1}{2}\theta_{1,2}V_o\left(\frac{\Lambda_{1,1}^{(2,2)} + \Lambda_{1,1}^{(1,1)}}{\Lambda_{1,1}^{(2,2)} - \Lambda_{1,1}^{(1,1)}}\right), \quad (45)$$

$$-\left[\Lambda_{1,2}^{(2,1)} + \frac{\theta_{2,1}^2 V_o^2}{2\left(\Lambda_{1,1}^{(2,2)} - \Lambda_{1,1}^{(1,1)}\right)}\right]\rho_{1,2}^{(1)} + \left[i\left(\Omega + \epsilon_{2,1}\right) - \Lambda_{1,2}^{(1,2)} + \frac{|\theta_{1,2}|^2 V_o^2}{2\left(\Lambda_{1,1}^{(2,2)} - \Lambda_{1,1}^{(1,1)}\right)}\right]\rho_{2,1}^{(1)}$$
$$+ \frac{\theta_{2,1}^2 V_o^2}{2\left(\Lambda_{1,1}^{(2,2)} - \Lambda_{1,1}^{(1,1)}\right)}\rho_{1,2}^{(-1)} - \frac{|\theta_{1,2}|^2 V_o^2}{2\left(\Lambda_{1,1}^{(2,2)} - \Lambda_{1,1}^{(1,1)}\right)}\rho_{2,1}^{(-1)} \cong \frac{1}{2}\theta_{2,1}V_o\left(\frac{\Lambda_{1,1}^{(2,2)} + \Lambda_{1,1}^{(1,1)}}{\Lambda_{1,1}^{(2,2)} - \Lambda_{1,1}^{(1,1)}}\right), \quad (46)$$

$$-\frac{|\theta_{1,2}|^2 V_o^2}{2\left(\Lambda_{1,1}^{(2,2)} - \Lambda_{1,1}^{(1,1)}\right)}\rho_{1,2}^{(1)} + \frac{\theta_{1,2}^2 V_o^2}{2\left(\Lambda_{1,1}^{(2,2)} - \Lambda_{1,1}^{(1,1)}\right)}\rho_{2,1}^{(1)}$$
$$+ \left[-i\left(\Omega + \epsilon_{2,1}\right) - \Lambda_{1,2}^{(1,2)} + \frac{|\theta_{1,2}|^2 V_o^2}{2\left(\Lambda_{1,1}^{(2,2)} - \Lambda_{1,1}^{(1,1)}\right)}\right]\rho_{1,2}^{(-1)} - \left[\Lambda_{1,2}^{(2,1)} + \frac{\theta_{1,2}^2 V_o^2}{2\left(\Lambda_{1,1}^{(2,2)} - \Lambda_{1,1}^{(1,1)}\right)}\right]\rho_{2,1}^{(-1)}$$
$$\cong \frac{1}{2}\theta_{1,2}V_o\left(\frac{\Lambda_{1,1}^{(2,2)} + \Lambda_{1,1}^{(1,1)}}{\Lambda_{1,1}^{(2,2)} - \Lambda_{1,1}^{(1,1)}}\right), \quad (47)$$

$$-\left[\Lambda_{1,2}^{(2,1)} + \frac{\theta_{2,1}^2 V_o^2}{2\left(\Lambda_{1,1}^{(2,2)} - \Lambda_{1,1}^{(1,1)}\right)}\right]\rho_{1,2}^{(-1)} + \left[-i\left(\Omega - \epsilon_{2,1}\right) - \Lambda_{1,2}^{(1,2)} + \frac{|\theta_{1,2}|^2 V_o^2}{2\left(\Lambda_{1,1}^{(2,2)} - \Lambda_{1,1}^{(1,1)}\right)}\right]\rho_{2,1}^{(-1)}$$
$$+ \frac{\theta_{2,1}^2 V_o^2}{2\left(\Lambda_{1,1}^{(2,2)} - \Lambda_{1,1}^{(1,1)}\right)}\rho_{1,2}^{(1)} - \frac{|\theta_{1,2}|^2 V_o^2}{2\left(\Lambda_{1,1}^{(2,2)} - \Lambda_{1,1}^{(1,1)}\right)}\rho_{2,1}^{(1)} \cong -\frac{1}{2}\theta_{2,1}V_o\left(\frac{\Lambda_{1,1}^{(2,2)} + \Lambda_{1,1}^{(1,1)}}{\Lambda_{1,1}^{(2,2)} - \Lambda_{1,1}^{(1,1)}}\right). \quad (48)$$

Multiplying (45) by $\theta_{2,1}$ and (46) by $\theta_{1,2}$, then adding the resulting expressions, gives

$$\rho_{2,1}^{(1)} \cong -\left[\frac{i\theta_{2,1}(\Omega - \epsilon_{2,1}) - \theta_{1,2}\Lambda_{1,2}^{(2,1)} - \theta_{2,1}\Lambda_{1,2}^{(1,2)}}{i\theta_{1,2}(\Omega + \epsilon_{2,1}) - \theta_{1,2}\Lambda_{1,2}^{(1,2)} - \theta_{2,1}\Lambda_{1,2}^{(2,1)}}\right]\rho_{1,2}^{(1)}. \quad (49)$$

Similarly, multiplying (47) by $\theta_{2,1}$ and (48) by $\theta_{1,2}$, then adding the resulting expressions, yields

$$\rho_{2,1}^{(-1)} \cong -\left[\frac{i\theta_{2,1}(\Omega + \epsilon_{2,1}) + \theta_{1,2}\Lambda_{1,2}^{(2,1)} + \theta_{2,1}\Lambda_{1,2}^{(1,2)}}{i\theta_{1,2}(\Omega - \epsilon_{2,1}) + \theta_{1,2}\Lambda_{1,2}^{(1,2)} + \theta_{2,1}\Lambda_{1,2}^{(2,1)}}\right]\rho_{1,2}^{(-1)}. \quad (50)$$

Substituting (49) into (45) and (50) into (47) results in two coupled equations involving the remaining coefficients, $\rho_{1,2}^{(\pm 1)}$, viz.

$$\left[i(\Omega - \epsilon_{2,1}) - \Lambda_{1,2}^{(-)}(\Omega) + \gamma_{1,2}^{(-)}(\Omega)\right]\rho_{1,2}^{(1)} - \gamma_{1,2}^{(+)}(\Omega)\,\rho_{1,2}^{(-1)} \cong -\frac{1}{2}\theta_{1,2}V_o\left(\frac{\Lambda_{1,1}^{(2,2)} + \Lambda_{1,1}^{(1,1)}}{\Lambda_{1,1}^{(2,2)} - \Lambda_{1,1}^{(1,1)}}\right), \quad (51)$$

$$\gamma_{1,2}^{(-)}(\Omega)\,\rho_{1,2}^{(1)} + \left[i(\Omega + \epsilon_{2,1}) + \Lambda_{1,2}^{(+)}(\Omega) - \gamma_{1,2}^{(+)}(\Omega)\right]\rho_{1,2}^{(-1)} \cong -\frac{1}{2}\theta_{1,2}V_o\left(\frac{\Lambda_{1,1}^{(2,2)} + \Lambda_{1,1}^{(1,1)}}{\Lambda_{1,1}^{(2,2)} - \Lambda_{1,1}^{(1,1)}}\right), \quad (52)$$

where we have defined

$$\Lambda_{1,2}^{(\pm)}(\Omega) = \Lambda_{1,2}^{(1,2)} - \Lambda_{1,2}^{(2,1)}\left[\frac{i\theta_{2,1}(\Omega \pm \epsilon_{2,1}) \pm \theta_{1,2}\Lambda_{1,2}^{(2,1)} \pm \theta_{2,1}\Lambda_{1,2}^{(1,2)}}{i\theta_{1,2}(\Omega \mp \epsilon_{2,1}) \pm \theta_{1,2}\Lambda_{1,2}^{(1,2)} \pm \theta_{2,1}\Lambda_{1,2}^{(2,1)}}\right], \quad (53)$$

$$\gamma_{1,2}^{(\pm)}(\Omega) = \frac{V_o^2}{2\left(\Lambda_{1,1}^{(2,2)} - \Lambda_{1,1}^{(1,1)}\right)}\left\{|\theta_{1,2}|^2 + \theta_{1,2}^2\left[\frac{i\theta_{2,1}(\Omega \pm \epsilon_{2,1}) \pm \theta_{1,2}\Lambda_{1,2}^{(2,1)} \pm \theta_{2,1}\Lambda_{1,2}^{(1,2)}}{i\theta_{1,2}(\Omega \mp \epsilon_{2,1}) \pm \theta_{1,2}\Lambda_{1,2}^{(1,2)} \pm \theta_{2,1}\Lambda_{1,2}^{(2,1)}}\right]\right\}. \quad (54)$$

Solving for $\rho_{1,2}^{(\pm 1)}$, we find

$$\rho_{1,2}^{(\pm 1)} \cong -\frac{\theta_{1,2}V_o}{2D_{1,2}(\Omega)}\left(\frac{\Lambda_{1,1}^{(2,2)} + \Lambda_{1,1}^{(1,1)}}{\Lambda_{1,1}^{(2,2)} - \Lambda_{1,1}^{(1,1)}}\right)\left[i(\Omega \pm \epsilon_{2,1}) \pm \Lambda_{1,2}^{(\pm)}(\Omega)\right], \quad (55)$$

with characteristic denominator

$$D_{1,2}(\Omega) = \left[i(\Omega - \epsilon_{2,1}) - \Lambda_{1,2}^{(-)}(\Omega)\right]\left[i(\Omega + \epsilon_{2,1}) + \Lambda_{1,2}^{(+)}(\Omega)\right]$$
$$- \gamma_{1,2}^{(+)}(\Omega)\left[i(\Omega - \epsilon_{2,1}) - \Lambda_{1,2}^{(-)}(\Omega)\right] + \gamma_{1,2}^{(-)}(\Omega)\left[i(\Omega + \epsilon_{2,1}) + \Lambda_{1,2}^{(+)}(\Omega)\right]. \quad (56)$$

Summarizing our analysis to this point, off-diagonal density matrix elements are purely oscillatory, with (38) and (55) implying

$$\rho_{1,2}(t) \cong -\frac{\theta_{1,2}V_o}{2D_{1,2}(\Omega)}\left(\frac{\Lambda_{1,1}^{(2,2)} + \Lambda_{1,1}^{(1,1)}}{\Lambda_{1,1}^{(2,2)} - \Lambda_{1,1}^{(1,1)}}\right)$$
$$\times \left\{\left[2i\Omega + \Lambda_{1,2}^{(+)}(\Omega) - \Lambda_{1,2}^{(-)}(\Omega)\right]\cos(\Omega t) + \left[-2\epsilon_{2,1} + i\Lambda_{1,2}^{(+)}(\Omega) + i\Lambda_{1,2}^{(-)}(\Omega)\right]\sin(\Omega t)\right\}, \quad (57)$$

and $\rho_{2,1}(t) = \rho_{1,2}(t)^*$. Diagonal density matrix elements are $\rho_{1,1}(t) = 1 - \rho_{2,2}(t)$ and $\rho_{2,2}(t)$, where $\rho_{2,2}(t)$ approximates to the dc term in the Fourier-like expansion of (38), which is the steady-state asymptote as $t \to \infty$. Specifically, from (42) and (55), we obtain $\rho_{2,2}^{(0)}$, from which we arrive at

$$\rho_{2,2}(t) \cong \frac{-1}{2\left(\Lambda_{1,1}^{(2,2)} - \Lambda_{1,1}^{(1,1)}\right)}\left\{\Lambda_{1,1}^{(1,1)}\right.$$
$$\left. - \frac{|\theta_{1,2}|^2 V_o^2}{2D_{1,2}(\Omega)}\left(\frac{\Lambda_{1,1}^{(2,2)} + \Lambda_{1,1}^{(1,1)}}{\Lambda_{1,1}^{(2,2)} - \Lambda_{1,1}^{(1,1)}}\right)\left[2i\epsilon_{2,1} + \Lambda_{1,2}^{(+)}(\Omega) + \Lambda_{1,2}^{(-)}(\Omega)\right]\right\} + h.c. \quad (58)$$

11

## B. The Frequency Response Near the Rabi Resonance

The next step is to transform the characteristic denominator of (56) into an expression that more clearly indicates the underlying frequency response of the transmon-resonator system. After some algebra, making use of (53) and (54), a preliminary expression is

$$\left[i\theta_{1,2}(\Omega - \epsilon_{2,1}) + \theta_{1,2}\Lambda_{1,2}^{(1,2)} + \theta_{2,1}\Lambda_{1,2}^{(2,1)}\right]\left[i\theta_{1,2}(\Omega + \epsilon_{2,1}) - \theta_{1,2}\Lambda_{1,2}^{(1,2)} - \theta_{2,1}\Lambda_{1,2}^{(2,1)}\right]D_{1,2}(\Omega) =$$

$$\theta_{1,2}^2\Bigg(\left[\Omega^2 - \epsilon_{2,1}^2 + 2i\Lambda_{1,2}^{(1,2)}\Omega + \Lambda_{1,2}^{(1,2)^2} - \Lambda_{1,2}^{(2,1)^2}\right]\left[\Omega^2 - \epsilon_{2,1}^2 - 2i\Lambda_{1,2}^{(1,2)}\Omega - \Lambda_{1,2}^{(1,2)^2} + \Lambda_{1,2}^{(2,1)^2}\right]$$

$$+ \frac{V_o^2}{\Lambda_{1,1}^{(2,2)} - \Lambda_{1,1}^{(1,1)}}\bigg\{(\Omega^2 - \epsilon_{2,1}^2)\left[2|\theta_{1,2}|^2\Lambda_{1,2}^{(1,2)} + \left(\theta_{1,2}^2 + \theta_{2,1}^2\right)\Lambda_{1,2}^{(2,1)}\right]$$

$$+ 2i|\theta_{1,2}|^2\Omega\left[2i\Lambda_{1,2}^{(1,2)}\Omega + \Lambda_{1,2}^{(1,2)^2} - \Lambda_{1,2}^{(2,1)^2}\right]\bigg\}\Bigg). \quad (59)$$

Next, since $\Omega \approx \epsilon_{2,1}$, we can write $\Omega^2 - \epsilon_{2,1}^2 \cong 2\epsilon_{2,1}(\Omega - \epsilon_{2,1})$ and make this type of substitution into the right side of the above equation. In this way, we arrive at

$$\left[i\theta_{1,2}(\Omega - \epsilon_{2,1}) + \theta_{1,2}\Lambda_{1,2}^{(1,2)} + \theta_{2,1}\Lambda_{1,2}^{(2,1)}\right]\left[i\theta_{1,2}(\Omega + \epsilon_{2,1}) - \theta_{1,2}\Lambda_{1,2}^{(1,2)} - \theta_{2,1}\Lambda_{1,2}^{(2,1)}\right]D_{1,2}(\Omega) \cong$$

$$4\epsilon_{2,1}^2\theta_{1,2}^2\Bigg\{(\Omega - \epsilon_{2,1})^2 + \frac{V_o^2\left[2|\theta_{1,2}|^2\Lambda_{1,2}^{(1,2)} + \left(\theta_{1,2}^2 + \theta_{2,1}^2\right)\Lambda_{1,2}^{(2,1)}\right]}{2\epsilon_{2,1}\left(\Lambda_{1,1}^{(2,2)} - \Lambda_{1,1}^{(1,1)}\right)}(\Omega - \epsilon_{2,1})$$

$$- \left(\frac{|\theta_{1,2}|^2 V_o^2}{\Lambda_{1,1}^{(2,2)} - \Lambda_{1,1}^{(1,1)}} - \Lambda_{1,2}^{(1,2)}\right)\Lambda_{1,2}^{(1,2)} - \frac{1}{4\epsilon_{2,1}^2}\left(\Lambda_{1,2}^{(1,2)^2} - \Lambda_{1,2}^{(2,1)^2}\right)^2$$

$$+ \frac{i}{2\epsilon_{2,1}}\left(\frac{|\theta_{1,2}|^2 V_o^2}{\Lambda_{1,1}^{(2,2)} - \Lambda_{1,1}^{(1,1)}} - 2\Lambda_{1,2}^{(1,2)}\right)\left(\Lambda_{1,2}^{(1,2)^2} - \Lambda_{1,2}^{(2,1)^2}\right)\Bigg\}, \quad (60)$$

where we have aggregated terms of the same power of $\Omega - \epsilon_{2,1}$. Assuming $\epsilon_{2,1} \gg |\Lambda_{k,k'}^{(l,l')}|$, then, of the zero-order terms in $\Omega - \epsilon_{2,1}$, the most significant are the leading contributions, such that we may further approximate the denominator as

$$\left[i\theta_{1,2}(\Omega - \epsilon_{2,1}) + \theta_{1,2}\Lambda_{1,2}^{(1,2)} + \theta_{2,1}\Lambda_{1,2}^{(2,1)}\right]\left[i\theta_{1,2}(\Omega + \epsilon_{2,1}) - \theta_{1,2}\Lambda_{1,2}^{(1,2)} - \theta_{2,1}\Lambda_{1,2}^{(2,1)}\right]D_{1,2}(\Omega) \cong$$

$$4\epsilon_{2,1}^2\theta_{1,2}^2\Bigg\{(\Omega - \epsilon_{2,1})^2 + \frac{V_o^2\left[2|\theta_{1,2}|^2\Lambda_{1,2}^{(1,2)} + \left(\theta_{1,2}^2 + \theta_{2,1}^2\right)\Lambda_{1,2}^{(2,1)}\right]}{2\epsilon_{2,1}\left(\Lambda_{1,1}^{(2,2)} - \Lambda_{1,1}^{(1,1)}\right)}(\Omega - \epsilon_{2,1})$$

$$- \left(\frac{|\theta_{1,2}|^2 V_o^2}{\Lambda_{1,1}^{(2,2)} - \Lambda_{1,1}^{(1,1)}} - \Lambda_{1,2}^{(1,2)}\right)\Lambda_{1,2}^{(1,2)}\Bigg\}. \quad (61)$$

The frequency roots in $\Omega$, which make the right side of the above expression tend to zero, are

$$\Omega_\pm \cong \epsilon_{2,1} - \frac{V_o^2\left[2|\theta_{1,2}|^2\Lambda_{1,2}^{(1,2)} + \left(\theta_{1,2}^2 + \theta_{2,1}^2\right)\Lambda_{1,2}^{(2,1)}\right]}{4\epsilon_{2,1}\left(\Lambda_{1,1}^{(2,2)} - \Lambda_{1,1}^{(1,1)}\right)}$$

$$\pm \frac{1}{4\epsilon_{2,1}\left(\Lambda_{1,1}^{(2,2)} - \Lambda_{1,1}^{(1,1)}\right)}\left\{V_o^4\left[2|\theta_{1,2}|^2\Lambda_{1,2}^{(1,2)} + \left(\theta_{1,2}^2 + \theta_{2,1}^2\right)\Lambda_{1,2}^{(2,1)}\right]^2\right.$$

$$\left. + 16\epsilon_{2,1}^2\Lambda_{1,2}^{(1,2)}\left(\Lambda_{1,1}^{(2,2)} - \Lambda_{1,1}^{(1,1)}\right)\left[|\theta_{1,2}|^2V_o^2 - \Lambda_{1,2}^{(1,2)}\left(\Lambda_{1,1}^{(2,2)} - \Lambda_{1,1}^{(1,1)}\right)\right]\right\}^{1/2}, \quad (62)$$

such that we can write (61) as

$$\left[i\theta_{1,2}(\Omega - \epsilon_{2,1}) + \theta_{1,2}\Lambda_{1,2}^{(1,2)} + \theta_{2,1}\Lambda_{1,2}^{(2,1)}\right]\left[i\theta_{1,2}(\Omega + \epsilon_{2,1}) - \theta_{1,2}\Lambda_{1,2}^{(1,2)} - \theta_{2,1}\Lambda_{1,2}^{(2,1)}\right]D_{1,2}(\Omega) \cong$$

$$4\epsilon_{2,1}^2\theta_{1,2}^2(\Omega - \Omega_-)(\Omega - \Omega_+). \quad (63)$$

Applying (63) to (57) and (58), and replacing $\Omega$ in the numerators with $\epsilon_{1,2}$ because $\Omega \approx \epsilon_{2,1}$, we have

$$\rho_{1,2}(t) \cong -\frac{\theta_{1,2}V_o}{8\epsilon_{2,1}^2\theta_{1,2}^2(\Omega - \Omega_-)(\Omega - \Omega_+)}\left(\frac{\Lambda_{1,1}^{(2,2)} + \Lambda_{1,1}^{(1,1)}}{\Lambda_{1,1}^{(2,2)} - \Lambda_{1,1}^{(1,1)}}\right)$$

$$\times \left\{\left[2i\epsilon_{2,1}X_{1,2} - \Lambda_{1,2}^{(2,1)}\left(Y_{1,2}^{(+)} + Y_{1,2}^{(-)}\right)\right]\cos(\Omega t)\right.$$

$$\left. + i\left[2\left(i\epsilon_{2,1} + \Lambda_{1,2}^{(1,2)}\right)X_{1,2} - \Lambda_{1,2}^{(2,1)}\left(Y_{1,2}^{(+)} - Y_{1,2}^{(-)}\right)\right]\sin(\Omega t)\right\}, \quad (64)$$

$$\rho_{2,2}(t) \cong \frac{-1}{2\left(\Lambda_{1,1}^{(2,2)} - \Lambda_{1,1}^{(1,1)}\right)}\left\{\Lambda_{1,1}^{(1,1)} - \frac{|\theta_{1,2}|^2V_o^2}{8\epsilon_{2,1}^2\theta_{1,2}^2(\Omega - \Omega_-)(\Omega - \Omega_+)}\left(\frac{\Lambda_{1,1}^{(2,2)} + \Lambda_{1,1}^{(1,1)}}{\Lambda_{1,1}^{(2,2)} - \Lambda_{1,1}^{(1,1)}}\right)\right.$$

$$\left. \times \left[2\left(i\epsilon_{2,1} + \Lambda_{1,2}^{(1,2)}\right)X_{1,2} - \Lambda_{1,2}^{(2,1)}\left(Y_{1,2}^{(+)} - Y_{1,2}^{(-)}\right)\right]\right\} + h.c., \quad (65)$$

where we have defined coefficients

$$X_{1,2} = \left(2i\epsilon_{2,1}\theta_{1,2} - \theta_{1,2}\Lambda_{1,2}^{(1,2)} - \theta_{2,1}\Lambda_{1,2}^{(2,1)}\right)\left(\theta_{1,2}\Lambda_{1,2}^{(1,2)} + \theta_{2,1}\Lambda_{1,2}^{(2,1)}\right), \quad (66)$$

$$Y_{1,2}^{(+)} = \left(2i\epsilon_{2,1}\theta_{1,2} - \theta_{1,2}\Lambda_{1,2}^{(1,2)} - \theta_{2,1}\Lambda_{1,2}^{(2,1)}\right)\left(2i\epsilon_{2,1}\theta_{2,1} + \theta_{1,2}\Lambda_{1,2}^{(2,1)} + \theta_{2,1}\Lambda_{1,2}^{(1,2)}\right), \quad (67)$$

$$Y_{1,2}^{(-)} = \left(\theta_{1,2}\Lambda_{1,2}^{(2,1)} + \theta_{2,1}\Lambda_{1,2}^{(1,2)}\right)\left(\theta_{1,2}\Lambda_{1,2}^{(1,2)} + \theta_{2,1}\Lambda_{1,2}^{(2,1)}\right). \quad (68)$$

13

Again, with $\epsilon_{2,1} \gg |\Lambda_{k,k'}^{(l,l')}|$, the amplitudes $X_{1,2}$ and $Y_{1,2}^{(\pm)}$ in both (64) and (65) can be simplified, such that

$$\rho_{1,2}(t) \cong \frac{1}{2}\Lambda_{1,2}^{(1,2)}\left(\frac{\Lambda_{1,1}^{(2,2)} + \Lambda_{1,1}^{(1,1)}}{\Lambda_{1,1}^{(2,2)} - \Lambda_{1,1}^{(1,1)}}\right)\left\{\frac{\theta_{1,2}\,V_o\left[\cos(\Omega t) + i\sin(\Omega t)\right]}{(\Omega - \Omega_-)(\Omega - \Omega_+)}\right\}, \tag{69}$$

$$\rho_{2,2}(t) \cong \frac{-1}{\Lambda_{1,1}^{(2,2)} - \Lambda_{1,1}^{(1,1)}}\left\{\Lambda_{1,1}^{(1,1)} + \frac{1}{2}\Lambda_{1,2}^{(1,2)}\left(\frac{\Lambda_{1,1}^{(2,2)} + \Lambda_{1,1}^{(1,1)}}{\Lambda_{1,1}^{(2,2)} - \Lambda_{1,1}^{(1,1)}}\right)\left[\frac{|\theta_{1,2}|^2 V_o^2}{(\Omega - \Omega_-)(\Omega - \Omega_+)}\right]\right\}. \tag{70}$$

## C. The Half-Width At Half Maximum and Corresponding Decoherence Time

A last step is to simplify the frequency roots, $\Omega_\pm$ of (62), consistent with our earlier approximations. Specifically, we can write (62) in terms of real and imaginary parts as

$$\Omega_\pm \cong \tilde{\epsilon}_{2,1} \pm i\Delta_{2,1}, \tag{71}$$

where the energy difference $\hbar\epsilon_{2,1}$ is shifted to a value of $\hbar\tilde{\epsilon}_{2,1}$ by virtue of the applied voltage, with

$$\tilde{\epsilon}_{2,1} = \epsilon_{2,1} - \frac{|\theta_{1,2}|^2 V_o^2 \Lambda_{1,2}^{(0)}}{2\epsilon_{2,1}\left(\Lambda_{1,1}^{(2,2)} - \Lambda_{1,1}^{(1,1)}\right)} \;;\quad \Lambda_{1,2}^{(0)} = \Lambda_{1,2}^{(1,2)} + \left(\frac{\theta_{1,2}^2 + \theta_{2,1}^2}{2|\theta_{1,2}|^2}\right)\Lambda_{1,2}^{(2,1)}. \tag{72}$$

Also, there is a Lorentzian half-width at half-maximum given by

$$\Delta_{2,1} = \frac{\Lambda_{1,2}^{(0)}}{2\epsilon_{2,1}\left(\Lambda_{1,1}^{(2,2)} - \Lambda_{1,1}^{(1,1)}\right)}\left\{\frac{4\epsilon_{2,1}^2 \Lambda_{1,2}^{(1,2)}\left(\Lambda_{1,1}^{(2,2)} - \Lambda_{1,1}^{(1,1)}\right)}{\Lambda_{1,2}^{(0)\,2}}\left[\Lambda_{1,2}^{(1,2)}\left(\Lambda_{1,1}^{(2,2)} - \Lambda_{1,1}^{(1,1)}\right) - |\theta_{1,2}|^2 V_o^2\right]\right.$$

$$\left. - |\theta_{1,2}|^4 V_o^4\right\}^{1/2}. \tag{73}$$

Note that the argument of the square root of $\Delta_{2,1}$ can change sign with increasing $V_o$, which occurs at a critical voltage given by

$$V_c = \frac{\epsilon_{2,1}}{|\theta_{1,2}|\left|\Lambda_{1,2}^{(0)}\right|}\sqrt{-2\Lambda_{1,2}^{(1,2)}\left(\Lambda_{1,1}^{(2,2)} - \Lambda_{1,1}^{(1,1)}\right)\left[1 + \sqrt{1 + \left(\frac{\Lambda_{1,2}^{(0)}}{\epsilon_{2,1}}\right)^2}\right]}. \tag{74}$$

This voltage is typically of the order of $V_c \approx 1\ \mu\text{V}$, so for voltages of interest, say 10 nV or less, we have $V_o \ll V_c$. Also for voltages of interest, we have $|\theta_{1,2}|\,V_o/\epsilon_{2,1} \approx 10^{-4}$, such that

14

this contribution to the fourth power in the square root expression of $\Delta_{2,1}$ can be neglected to good approximation. This simplifies $\Delta_{2,1}$ to

$$\Delta_{2,1} \cong -\Lambda_{1,2}^{(1,2)} \sqrt{1 - \frac{|\theta_{1,2}|^2 V_o^2}{\Lambda_{1,2}^{(1,2)} \left(\Lambda_{1,1}^{(2,2)} - \Lambda_{1,1}^{(1,1)}\right)}} \; ; \quad V_o \ll V_c. \tag{75}$$

Since $\Lambda_{1,1}^{(2,2)} - \Lambda_{1,1}^{(1,1)} = -2\Lambda_{1,2}^{(1,2)} = 2|\Lambda_{1,2}^{(1,2)}|$, as can be demonstrated, we can write (75) as

$$\Delta_{2,1} \cong \frac{1}{\tau}\sqrt{1 + \frac{1}{2}|\theta_{1,2}|^2 V_o^2 \tau^2} \; ; \quad \tau = \frac{1}{\left|\Lambda_{1,2}^{(1,2)}\right|}, \tag{76}$$

where $\tau$ is a coherence time measurable from the observed frequency response. Now, the pertinent density matrix elements of (69) and (70) can be expressed as

$$\rho_{1,2}(t) \cong -\frac{1}{4}\left(\Lambda_{1,1}^{(1,1)} + \Lambda_{1,1}^{(2,2)}\right)\left[\frac{\theta_{1,2} V_o}{(\Omega - \tilde{\epsilon}_{2,1})^2 + \Delta_{2,1}^2}\right] e^{i\Omega t}, \tag{77}$$

$$\rho_{2,2}(t) \cong \frac{\tau}{2}\left\{\frac{1}{4}\left(\Lambda_{1,1}^{(1,1)} + \Lambda_{1,1}^{(2,2)}\right)\left[\frac{|\theta_{1,2}|^2 V_o^2}{(\Omega - \tilde{\epsilon}_{2,1})^2 + \Delta_{2,1}^2}\right] - \Lambda_{1,1}^{(1,1)}\right\}. \tag{78}$$

And, for the expectation value of the spin, as given by (34), we find $\langle \mathbf{S}(t) \rangle \cong -\rho_{2,2}(t)/2\,\hat{z}$, so

$$\langle \mathbf{S}(t) \rangle \cong \frac{\tau}{4}\left\{\Lambda_{1,1}^{(1,1)} - \frac{1}{4}\left(\Lambda_{1,1}^{(1,1)} + \Lambda_{1,1}^{(2,2)}\right)\left[\frac{|\theta_{1,2}|^2 V_o^2}{(\Omega - \tilde{\epsilon}_{2,1})^2 + \Delta_{2,1}^2}\right]\right\} \hat{z}, \tag{79}$$

as $t \to \infty$.

### ACKNOWLEDGMENTS

### Appendix A: Derivation of the Canonical Transformation

The quadratic terms of the second-quantized Hamiltonian of (11), given by

$$\mathscr{H}_o^{(0)} = \sum_{\mu=\pm} \hbar\omega_\mu \left(a_\mu^\dagger a_\mu + \frac{1}{2}\right) + \frac{1}{4}i\hbar\tilde{g}\left(a_+^\dagger + a_+\right)\left(a_-^\dagger - a_-\right), \tag{A1}$$

15

can be placed in diagonal form via a similarity transformation, $\mathcal{S}$, determined from the equations of motion of $a_\mu$ and $a_\mu^\dagger$. Specifically, with $A_{\mu,+} = a_\mu$ and $A_{\mu,-} = a_\mu^\dagger$, we can write the equations of motion of the quadratic limit in a matrix-element form as

$$i\hbar \frac{d}{dt} A_{\mu,m}(t) = \left[A_{\mu,m}(t), \mathcal{H}_o^{(0)}\right] = \sum_{\nu,m'} \mathcal{D}_{\mu,m;\nu,m'} A_{\nu,m'}(t), \tag{A2}$$

where indices $\mu, \nu = \pm$ and $m, m' = \pm$, and the $4 \times 4$ dynamical matrix, $\mathcal{D}$, has elements defined as

$$\mathcal{D}_{\mu,m;\mu,m'} = m\,\hbar\omega_\mu \delta_{m,m'}, \tag{A3}$$

$$\mathcal{D}_{-,m;+,m'} = \frac{1}{4} i\hbar\tilde{g}, \quad \mathcal{D}_{+,m;-,m'} = -mm'\mathcal{D}_{-,m;+,m'}. \tag{A4}$$

In the above, where indices $m, m' = \pm$ appear algebraically, they represent integers $\pm 1$, respectively.

Assuming a time evolution of ladder operators of the form $A_{\mu,m}(t) = A_{\mu,m} \exp(-i\omega t)$ we can construct the eigenproblem as

$$\sum_{\nu,m'} \mathcal{D}_{\mu,m;\nu,m'} A_{\nu,m'} = \hbar\omega\, A_{\mu,m}. \tag{A5}$$

Diagonalizing $\mathcal{D}$, whereby we introduce a new index $\sigma \in \{\uparrow, \downarrow\}$, we obtain four eigenvalues for $\omega$ that can be expressed as $\pm\omega_\uparrow$ and $\pm\omega_\downarrow$. With the algebraic meaning $\sigma = +1$ ($\sigma = -1$) when index $\sigma = \uparrow$ ($\sigma = \downarrow$), we can write $\omega_\uparrow$ and $\omega_\downarrow$ succinctly as

$$\omega_\sigma = \sqrt{\frac{1}{2}\left[\omega_+^2 + \omega_-^2 + \sigma\sqrt{(\omega_+^2 - \omega_-^2)^2 + \tilde{g}^2\,\omega_+\omega_-}\right]}. \tag{A6}$$

Since dynamical matrix $\mathcal{D}$ is non-Hermitian, similarity transformation $\mathcal{S}$ is formed from columns of right eigenvectors, $e$, while its inverse, $\mathcal{S}^{-1}$, is formed from rows of left eigenvectors, $u$. Respectively, introducing index $s = \pm$, we can define elements of the right and left eigenvectors as

$$e_{\sigma,s}(\mu,m) = \frac{1}{2}\xi_{\sigma,\mu}\left(\frac{\omega_\mu + ms\,\omega_\sigma}{\omega_\sigma}\right) e^{i\pi[\sigma m(1-\mu) - s(1-\sigma)]/4}, \quad u_{\sigma,s}(\mu,m) = ms\, e_{\sigma,s}(\mu,m)^*, \tag{A7}$$

where

$$\xi_{\sigma,\mu} = \sqrt{\frac{\omega_\sigma}{\omega_\mu}\left(\frac{\omega_{\bar{\mu}}^2 - \omega_\sigma^2}{\omega_{\bar{\sigma}}^2 - \omega_\sigma^2}\right)}, \tag{A8}$$

with the meaning $\bar{\mu} = -\mu$ and $\bar{\sigma} = \downarrow$ ($\bar{\sigma} = \uparrow$) when $\sigma = \uparrow$ ($\sigma = \downarrow$). Also, similar to $m = \pm$, the algebraic meaning of $s = +$ ($s = -$) is $+1$ ($-1$).

Due to the form of $\mathscr{D}$, the right and left eigenvectors satisfy the relations

$$e_{\sigma,-s}(\mu, -m) = e_{\sigma,s}(\mu, m)^*, \quad u_{\sigma,-s}(\mu, -m) = u_{\sigma,s}(\mu, m)^*, \tag{A9}$$

in addition to the definitions of completeness and orthonormality, viz.

$$\sum_{\sigma,s} e_{\sigma,s}(\mu; m)\, u_{\sigma,s}(\nu, m') = \delta_{\mu,\nu}\delta_{m,m'}, \quad \sum_{\mu,m} u_{\sigma,s}(\mu, m)\, e_{\sigma',s'}(\mu, m) = \delta_{\sigma,\sigma'}\delta_{s,s'}. \tag{A10}$$

In the canonical space we can define new ladder operators $a_\sigma$ and $a_\sigma^\dagger$, which satisfy the usual commutation relations. These can be expressed in jump-operator form by defining $A_{\sigma,+} = a_\sigma$ and $A_{\sigma,-} = a_\sigma^\dagger$, analogous to our previous definitions of $A_{\mu,+} = a_\mu$ and $A_{\mu,-} = a_\mu^\dagger$. Using the right and left eigenvectors, we then have

$$A_{\sigma,s} = \sum_{\mu,m} u_{\sigma,s}(\mu, m)\, A_{\mu,m}, \quad A_{\mu,m} = \sum_{\sigma,s} e_{\sigma,s}(\mu, m)\, A_{\sigma,s}. \tag{A11}$$

A consequence of the underlying Lie algebra of the ladder operators of second quantization is that, while $\mathcal{S}^{-1}\mathscr{D}\mathcal{S}$ renders the dynamical matrix diagonal, the operations $\mathcal{S}^\dagger \mathscr{H}_o^{(0)} \mathcal{S}$ simultaneously render the quadratic Hamiltonian diagonal [9]. We leave to the reader the verification of

$$\mathcal{S}^\dagger \mathscr{H}_o^{(0)} \mathcal{S} = \sum_{\sigma \in \{\uparrow, \downarrow\}} \hbar\omega_\sigma \left( a_\sigma^\dagger a_\sigma + \frac{1}{2} \right). \tag{A12}$$

---


\fi

\end{document}